\documentclass[sigconf]{acmart}

\usepackage{graphicx}
\usepackage{amsmath}
\usepackage{booktabs}
\usepackage{hyperref}
\usepackage{float}
\usepackage{subfigure}
\usepackage{caption} 
\usepackage{algorithm}
\usepackage{algpseudocode}
\usepackage{hyperref} 
\usepackage{colortbl }
\usepackage{subcaption}
\usepackage{gensymb}
\usepackage{threeparttable}
\AtBeginDocument{%
  }

\def \geoind{\texttt{Geo-Ind}}
\def \dataset{\texttt{Yjmob100k}} 

\copyrightyear{2026}
\acmYear{2026}
\setcopyright{cc}
\setcctype{by}
\acmConference[ASIA CCS '26]{ACM Asia Conference on Computer and Communications Security}{June 01--05, 2026}{Bangalore, India}
\acmBooktitle{ACM Asia Conference on Computer and Communications Security (ASIA CCS '26), June 01--05, 2026, Bangalore, India}
\acmDOI{10.1145/3779208.3805978}
\acmISBN{979-8-4007-2356-8/2026/06}

\begin{document}

\title{How Tough Is Location Anonymization? Re-identifying 100K Real-User Trajectories in Japan}
\sloppy



\author{Abhishek Kumar Mishra}
\affiliation{%
  \institution{Inria}
  \city{Lyon}
  \country{France}
}
\email{abhishek.mishra@inria.fr}

\author{Mathieu Cunche}
\affiliation{%
  \institution{INSA-Lyon, Inria, CITI, UR3720}
  \city{Villeurbanne}
  \country{France}
}
\email{mathieu.cunche@insa-lyon.fr}






\author{Héber H. Arcolezi}

\affiliation{
  \institution{Inria, Grenoble, France}
  \city{}
  \country{}
}

\affiliation{
  \institution{ÉTS Montréal, Montréal, Canada}
  \city{}
  \country{}
}

\email{heber.hwang-arcolezi@etsmtl.ca}

\renewcommand{\shortauthors}{Mishra, Cunche, \& Arcolezi}

\begin{abstract}
Mobility traces are among the most revealing forms of personal data, yet trajectory releases are often protected only by ad hoc transformations. We stress-test such practices on recently-released \texttt{YJMob100K}, an anonymized dataset of 100{,}000 user trajectories in Japan. First, we show that the applied protection leaves enough spatial and temporal structure to recover both the real-world geographic frame and the actual calendar timeline by exploiting density signatures, urban correlations, and temporal activity profiles. On top of this reconstruction, we quantify privacy risks through trajectory-level metrics that capture spatio-temporal $k$-anonymity, $m$-point unicity, home--work and multi-anchor uniqueness, and exposure to secluded and sensitive locations. These metrics reveal extensive re-identification surfaces: a small number of observations, anchors, or sensitive venues often suffices to uniquely pinpoint users or their social neighborhoods. Finally, we evaluate representative sanitization strategies: geo-indistinguishability, local differential privacy, and aggressive spatial de-structuring; and observe a consistent pattern: strong privacy parameters destroy downstream utility, while utility-preserving settings leave structural leakage largely intact. Overall, our findings show that current sanitization techniques are insufficient for large-scale mobility data, and they highlight the urgent need for trajectory-aware privacy mechanisms and stronger publication standards.
\end{abstract}



\maketitle

\section{Introduction}

Mobility data is central to modern urban analytics, transportation planning, and epidemiological modeling. At the same time, human movement is highly identifying: trajectories encode stable routines, anchor locations, and social structure~\cite{gonzalez2008understanding}. Even after direct identifiers are removed, an adversary can often link pseudonymous traces to external information and re-identify individuals~\cite{montjoye_unique_2013}. This tension becomes particularly acute for large-scale, high-resolution releases, where rich structure is what makes the data useful. 

In this paper, we take a close look at \texttt{YJMob100K}~\cite{yabe2024YJMob100K}, a recently-published public mobility dataset containing GPS trajectories for 100{,}000 users in Japan. The release replaces absolute timestamps and geographic coordinates with values defined relative to unknown spatial and temporal reference points, with the stated goal of breaking the linkage to real locations and dates. However, the transformed trajectories still preserve dense structural signals: spatial density gradients, road-aligned corridors, recurrent commuting pathways, and weekly temporal rhythms; that act as latent ``privacy beacons'' at the national scale.

We build an adversarial pipeline that operates directly on these structural cues. While prior work has hinted at the region of origin~\cite{pinter_revealing_2024}, we go significantly further by recovering meter-level coordinates, exact calendar dates, and conducting a full sensitivity analysis of potential defenses. Specifically, we first use population-density signatures and urban-form correlations to match \texttt{YJMob100K} to real cities and recover the hidden geographic frame. We then refine this alignment to the meter level, mapping anonymized grid cells back to latitude-longitude coordinates. In parallel, we exploit temporal activity profiles, weekday/weekend patterns, public holidays, and event-specific spikes to re-identify the true calendar timeline. Together, these steps convert the ``anonymized'' release into a time-stamped, geo-referenced dataset of real users moving through a real Japanese city.

On top of this re-identification, we systematically quantify privacy risks using state-of-the-art, trajectory-level metrics. Our framework covers spatio-temporal $k$-anonymity under realistic adversary queries, $m$-point unicity of partial trajectories, home--work and multi-anchor uniqueness, and exposure to secluded and sensitive locations. These metrics reveal extensive attack surfaces: small sets of points, stable anchors, or rare venues often suffice to single out users or small social neighborhoods.

Finally, we evaluate three representative sanitization strategies that span widely adopted and frequently advocated approaches in location privacy.
First, we consider \emph{geo-indistinguishability}~\cite{Andrs2013}, which is widely regarded as a state-of-the-art formal privacy guarantee for protecting individual locations in the absence of a trusted curator.
Second, we study \emph{Generalized Randomized Response (GRR)}~\cite{kairouz2016discrete}, a well-established primitive in the local differential privacy (LDP) literature that has been used for discrete location reporting and large-scale aggregate mobility analysis~\cite{Quercia2011spotme}.
Third, we examine an \emph{aggressive spatial de-structuring baseline} that deliberately destroys global spatial coherence while preserving marginal density distributions, serving as an upper bound on defenses that rely on structure removal rather than point-wise noise.
Across all three mechanisms, we observe a consistent pattern: configurations that preserve analytical utility leave substantial structural leakage intact, whereas parameter regimes that provide strong privacy guarantees severely distort population distributions, degrade anchor inference, and compromise downstream analytics.

In summary, our main contributions are:
\begin{enumerate}
    \item We develop a scalable spatial re-identification methodology, 
    enabling recovery of the hidden urban area in \texttt{YJMob100K} and refinement to meter-level alignment between grid cells and real-world coordinates.

    \item We re-identify the dataset’s true calendar 
    , thereby restoring real-world timestamps for 100{,}000 users.

    \item We propose a comprehensive set of trajectory-level privacy metrics quantifying identity \& attribute leakage 
    once structural re-identification is possible.

    \item We conduct an empirical evaluation of representative sanitization strategies on the re-identified dataset, revealing a pronounced privacy-utility trade-off and showing that current mechanisms fall short for large-scale mobility releases.
\end{enumerate}

Our findings highlight a serious and underappreciated risk: once structural information in space and time is exposed, national-scale mobility releases can be pulled back into the real world with high fidelity, enabling detailed profiling of real individuals and communities. 
\textit{To our knowledge, this work provides the first end-to-end reversal of a modern large-scale anonymized mobility release under realistic deployment assumptions.} The code is available at \url{https://github.com/miishra/Re-identifying_100k_Trajectories}. 

\section{Related Work}

Pinter et al.~\cite{pinter_revealing_2024} analyze the \texttt{YJMob100K} dataset and demonstrate that its anonymization preserves sufficient spatial structure to enable identification of the underlying geographic region via density-based matching.
Their analysis provides early evidence that relative-coordinate releases can leak large-scale geographic information.
However, they do not recover fine-grained spatial coordinates, do not address temporal re-identification, and do not study how alternative sanitization strategies from the literature would behave on \texttt{YJMob100K}.
As a result, their work stops short of quantifying downstream privacy risks or translating these findings into actionable guidance for dataset publication.

A substantial body of prior work identifies location data as among the most privacy-sensitive forms of personal information.
Location traces enable inference of sensitive personal attributes~\cite{zhong_you_2015,drakonakis_please_2019} and frequently allow adversaries to single out individuals from large populations~\cite{montjoye_unique_2013}.
Crucially, this identifiability arises from structural properties of mobility data, such as regularity, sparsity, and anchor locations, rather than from explicit identifiers, making anonymization particularly challenging.

The literature on privacy-preserving data publishing proposes a wide range of mitigation strategies~\cite{fung2010privacy}.
While general-purpose anonymization techniques are relatively well understood for tabular data, extending these approaches to spatial data~\cite{acs_case_2014,Quercia2011spotme} and mobility traces remains difficult~\cite{fiore2019privacy}.
The sequential, highly correlated nature of trajectories complicates the application of both syntactic anonymization and formal privacy guarantees~\cite{Primault2019,fiore2019privacy,MirandaPascual2023}.
Empirical studies consistently show that anonymized mobility datasets remain vulnerable to a variety of re-identification and inference attacks, even after substantial perturbation~\cite{douriez_anonymizing_2016,Xu2017Trajectory,Pyrgelis2017What,Gambs2010,pyrgelis2017knock,gambs2014anonymization}, highlighting a persistent privacy-utility trade-off. 

\begin{figure}[t!]
    \centering
    \subfigure
    {\includegraphics[width=0.98\linewidth]{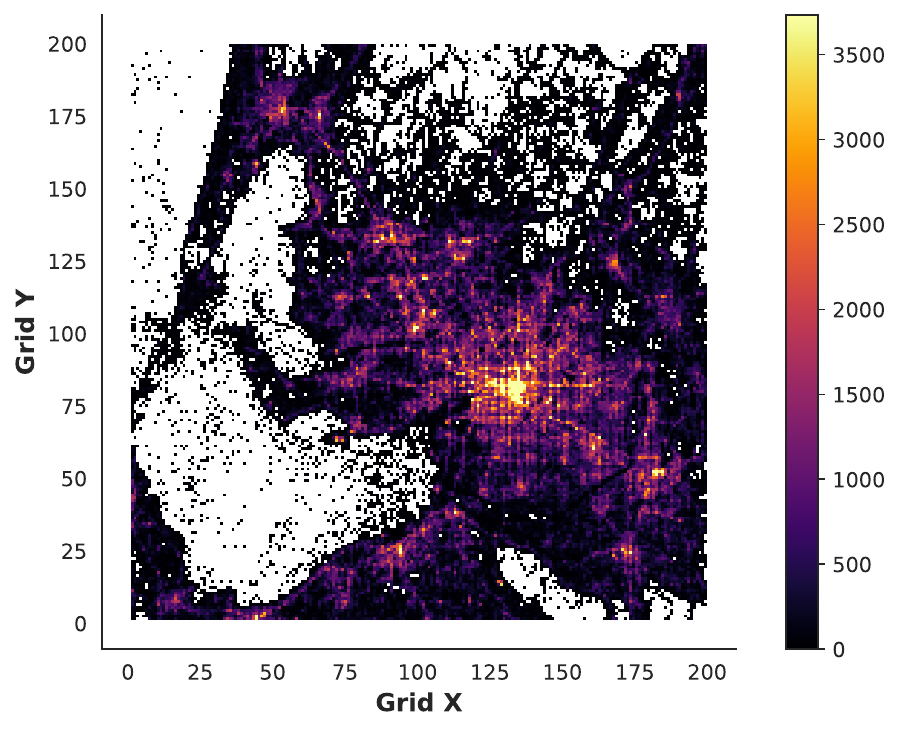} \label{fig:raw_trajectories}}
    \vspace{-0.45cm}
    \caption{Population densities of the released dataset. Grid-based trajectories in \texttt{YJMob100K} preserve structured mobility that aligns with urban infrastructure.}
\end{figure}

This work extends prior findings by moving from isolated vulnerability demonstrations to an end-to-end, adversarial evaluation of a real-world mobility release. 
Whereas earlier studies primarily document identifiability or inference risks in abstract or controlled settings, we jointly reconstruct both spatial and temporal reference and use this reconstruction as a basis for systematic privacy measurement. 
By evaluating representative sanitization mechanisms under realistic utility constraints, our study connects structural re-identification directly to downstream privacy-utility trade-offs, offering an empirical perspective on how widely used defenses behave when confronted with population-scale mobility structure.

\section{\texttt{YJMob100K} Dataset}

The \texttt{YJMob100K} dataset comprises GPS trajectories from 100{,}000 anonymized users across Japan. Each record includes time-stamped latitude-longitude coordinates, a user identifier, and additional metadata. The dataset also provides anonymized Points of Interest (PoIs), which represent frequently visited user locations.

Figure~\ref{fig:raw_trajectories} provides a spatial overview of the grid-based movement trajectories. The patterns show structured mobility in an urban region: dense clusters, directional flows, and corridor-like traces that track transportation infrastructure and metropolitan dynamics. Even after anonymization, the retained spatial organization, including recurrent pathways and grid connectivity, mirrors real geographies such as transit axes and high-density cores. Prominent clusters suggest the presence of landmarks or PoIs that later act as strong anchors for re-identification. 

\begin{figure}[t!]
    \centering
    \subfigure[Map of Nagoya]
    {\includegraphics[width=0.91\linewidth]{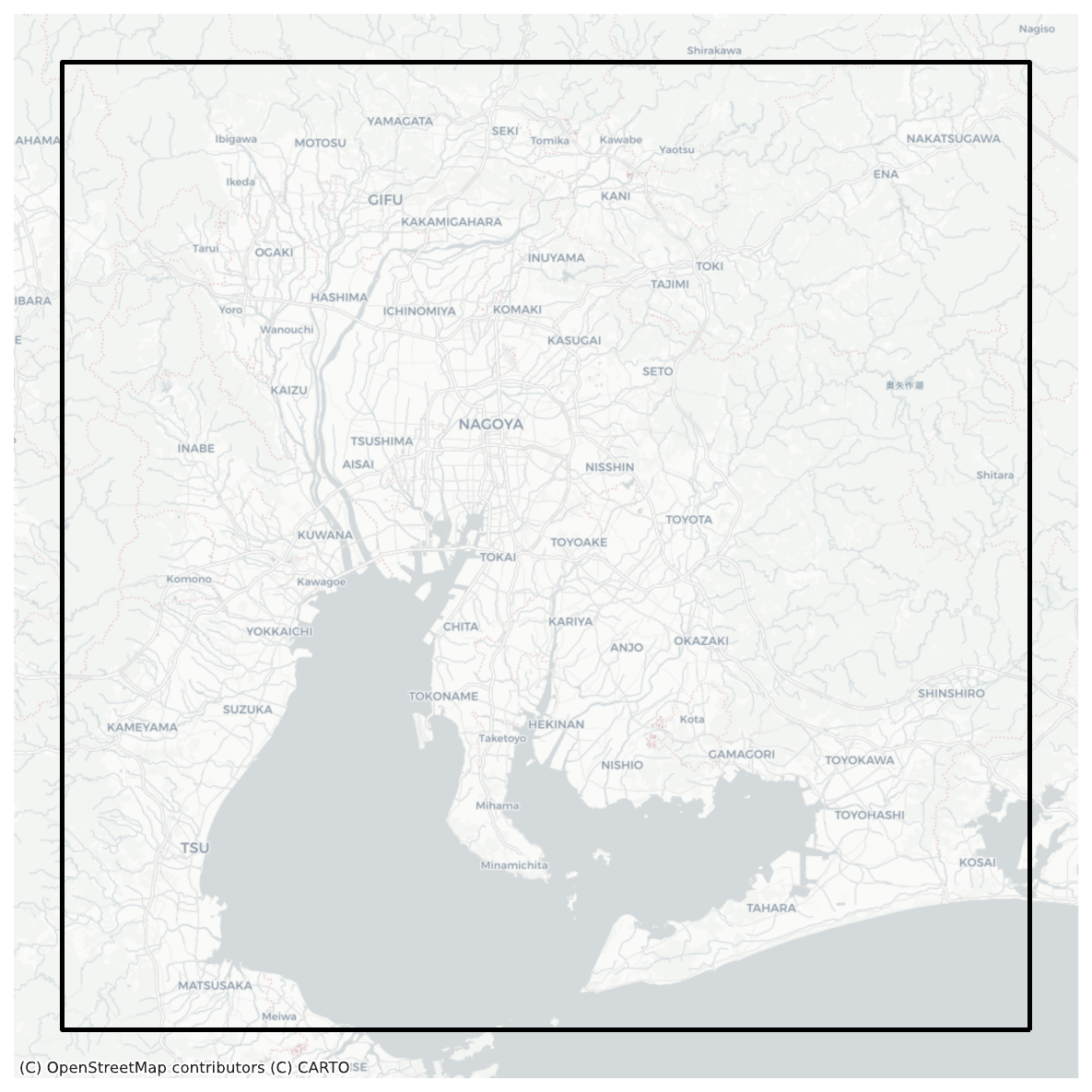} \label{fig:nagoya_map}}
    \hfill
    \vspace{-0.4cm}
    \subfigure[Re-identified dataset]
    {\includegraphics[width=0.95\linewidth]{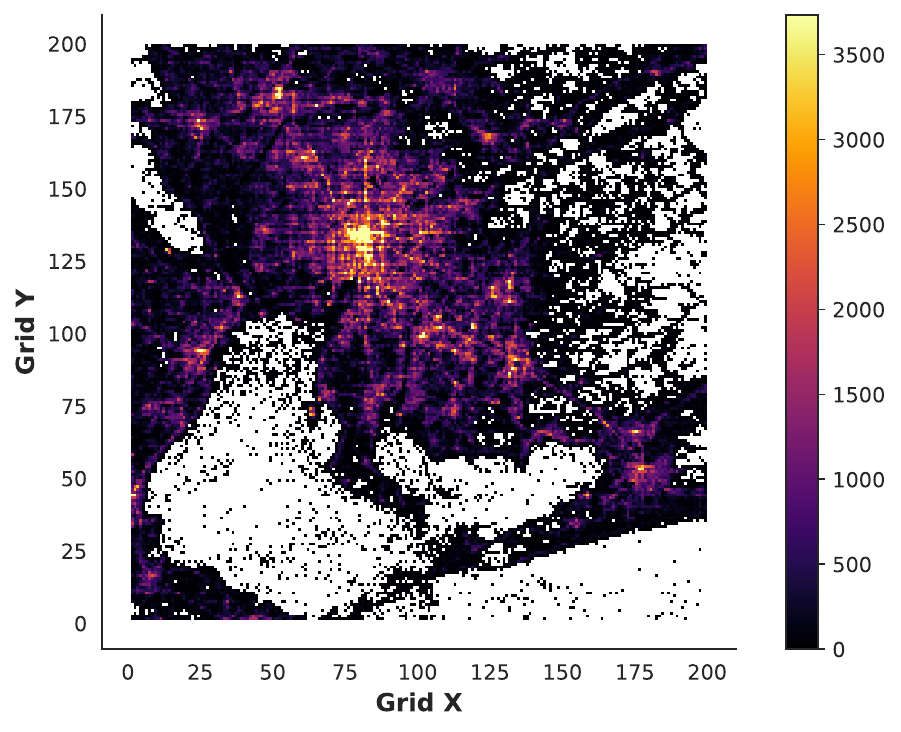} \label{fig:transformed_trajectories}}
    \vspace{-0.45cm}
    \caption{Population densities of the re-identified dataset. (a) Map of Nagoya. (b) The same traces after our re-identification and alignment procedure (Algorithm~\ref{alg:re-identification}), which reveals the underlying city and geography.}
\end{figure}

\section{Re-Identifying the Spatial Dimension}
\label{sec:reidentification-spatial}

We re-identify the real-world location of \texttt{YJMob100K} using spatial structure and density. Our core insight is simple and powerful: an anonymization pipeline that preserves density and connectivity also preserves the city-level ``fingerprint'' of the underlying geography. We exploit that fingerprint with public population datasets.

\subsection{Density-Based Re-identification Metric}

We test the hypothesis that the anonymized dataset retains structural properties that match real population distributions. For \texttt{YJMob100K}, we compute daily spatial density as:
\begin{equation}
    D(d) = \sum_{g \in G} P_g(d),
\end{equation}
where \(D(d)\) denotes the aggregated density on day \(d\), \(P_g(d)\) denotes the point count in grid cell \(g\), and \(G\) denotes the set of spatially selected grid cells. For comparison, we use public population density data~\cite{japan_data} for Japan's ten largest cities. These distributions exhibit stable urban structure, including dense cores, corridor-like gradients, and tapering suburban regions. A successful anonymization pipeline must break the linkage between those structure markers and the released trajectory density.

\subsection{Re-Identification of Location}

To infer the origin location of the dataset, we apply the re-identification procedure defined in Algorithm~\ref{alg:re-identification}. The algorithm takes as input the anonymized dataset $D_{\text{raw}}$, a set of public population datasets $D_{\text{public}}$ corresponding to Japan’s ten largest cities, and a transformation set $T$. The goal is to identify the most likely city of origin $C^*$ and the transformation $T^*$ that maximizes spatial similarity, with all intermediate correlation values stored in $S$.

We define a set of geometric transformations that account for potential obfuscation applied to $D_{\text{raw}}$: horizontal flip $F_x(D)$, vertical flip $F_y(D)$, 90$^\circ$ clockwise rotation $R_{90}(D)$, and 90$^\circ$ counterclockwise rotation $R_{-90}(D)$. Combining these operations produces eight transformed variants, which form the transformation set $T$. 

For each transformed dataset $T_i \in T$, we compute a Spearman correlation $C_i(j)$ between $T_i(D_{\text{raw}})$ and each public dataset $D_{\text{public}}(j)$, where $j \in \{1, \dots, 10\}$. To ensure robustness, we compute correlation over clustered spatial regions: we divide the full $200 \times 200$ grid into 25 non-overlapping clusters of size $40 \times 40$. We store the resulting scores in the dictionary $S$.

We identify the optimal match by selecting $(C^*, T^*) = \arg \max_{(T_i, j)} C_i(j)$. The final output includes the re-identified city $C^*$ and the optimal transformation $T^*$.

\begin{algorithm}[t!]
\caption{Re-Identification Algorithm}
\label{alg:re-identification}
\begin{algorithmic}[1]
\State \textbf{Input:} $D_{\text{raw}}$, $D_{\text{public}}$
\State \textbf{Output:} $C^*$, $T^*$, $S$
\State $F_x(D) = \{ (x, -y) \mid (x, y) \in D \}$ \Comment{Flip x-axis}
\State $F_y(D) = \{ (-x, y) \mid (x, y) \in D \}$ \Comment{Flip y-axis}
\State $R_{90}(D) = \{ (y, -x) \mid (x, y) \in D \}$ \Comment{Rotate $+90^\circ$}
\State $R_{-90}(D) = \{ (-y, x) \mid (x, y) \in D \}$ \Comment{Rotate $-90^\circ$}
\State $T = \{D, F_x(D), F_y(D), F_x(F_y(D)), R_{90}(D), \newline R_{-90}(D), R_{90}(F_x(D)), R_{90}(F_y(D))\}$
\State $S = \{\}$
\For{$T_i \in T$}
    \For{$j = 1$ \textbf{to} $10$}
        \State $C_i(j) = C_{\text{corr}}(T_i(D_{\text{raw}}), D_{\text{public}}(j))$
        \State $S[(T_i, j)] = C_i(j)$
    \EndFor
\EndFor
\State $(C^*, T^*) = \arg \max_{(T_i, j) \in S} C_i(j)$
\State \textbf{Return} $(C^*, T^*, S)$
\end{algorithmic}
\end{algorithm} 

\begin{figure}[t!]
    \centering
    \subfigure[Individual grid correlation]{\includegraphics[width=0.98\linewidth]{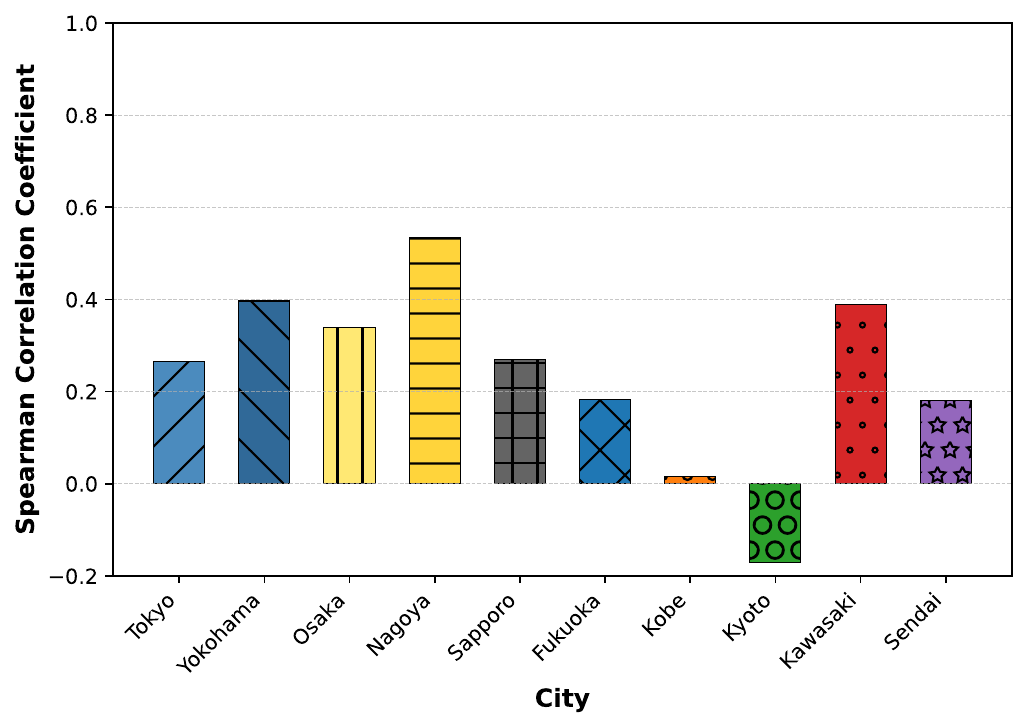} \label{fig:grid_correlation}}
    \hfill
    \vspace{-0.45cm}
    \subfigure[Cluster-aggregated correlation]{\includegraphics[width=0.98\linewidth]{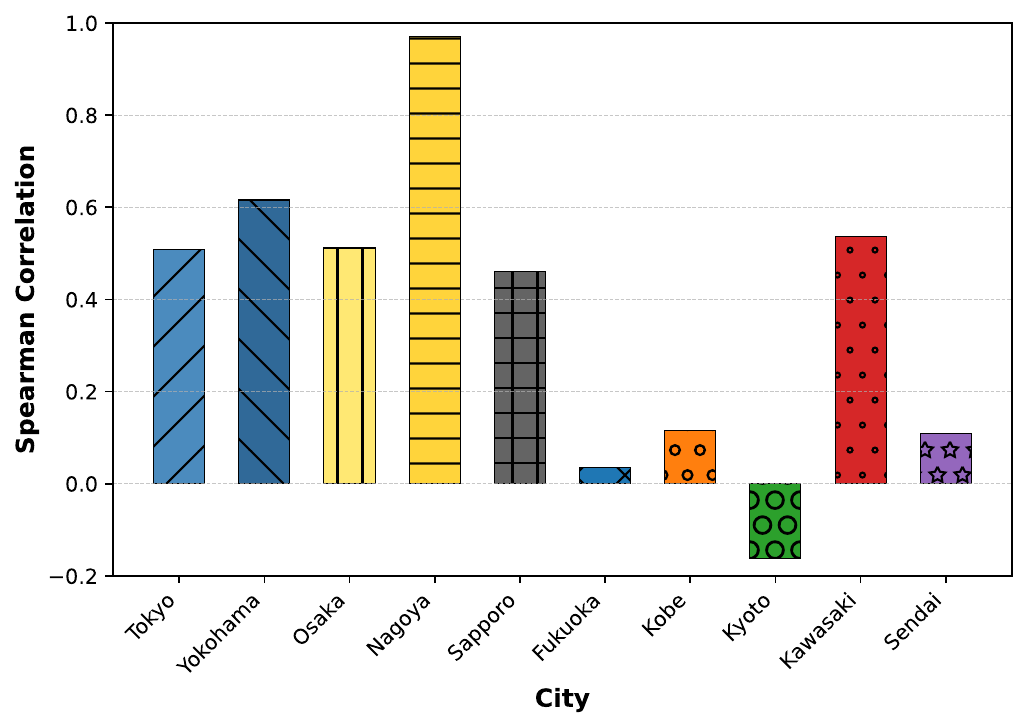} \label{fig:clustered_correlation}}
    \vspace{-0.45cm}
    \caption{Spearman correlations between the anonymized dataset and population data across 10 major Japanese cities. Nagoya emerges as the dominant match.}
    \label{fig:correlation_comparison}
\end{figure}

Figure~\ref{fig:transformed_trajectories} visualizes the anonymized dataset after applying the optimal transformation (\(R_{90}(F_y(D))\)), and Figure~\ref{fig:nagoya_map} shows the corresponding match with the city of Nagoya. These transformations obscure orientation but preserve intrinsic movement structure, including cluster layout and displacement ranges. By inverting the transformations, we recover alignment with real geographies.

Quantitatively, Figure~\ref{fig:grid_correlation} and Figure~\ref{fig:clustered_correlation} show Spearman correlations between the transformed dataset and public population data. Nagoya consistently yields the strongest match, and the clustered strategy boosts robustness by suppressing cell-level noise. Under clustering, the correlation for Nagoya approaches unity, which indicates a near-perfect structural match.

\subsection{Fine-Grained Spatial Re-identification}

After identifying Nagoya as the most probable city, we refine alignment with a fine-tuning step using hill climbing over a 100 km $\times$ 100 km area centered around Nagoya. This step targets exact latitude-longitude alignment that maximizes Spearman correlation between the public population grid and the anonymized dataset.

Starting from $(35.05, 136.96)$, the algorithm iteratively perturbs the center by small steps (0.01$^{\circ}$) and retains configurations that increase correlation. The process converges at $(35.055019, 136.971202)$ once further perturbations yield no improvement. This procedure recovers the geocoordinates corresponding to anonymized grid cells in \texttt{YJMob100K} and completes spatial re-identification with meter-level precision. The recovered alignment, together with clustered correlation near 1.0 for Nagoya (Figure~\ref{fig:correlation_comparison}), confirms reliable spatial recovery. Prior work~\cite{pinter_revealing_2024} hints at Nagoya as the underlying city; our method completes the attack surface by uncovering exact locations and validating the match through fine-grained optimization.

\section{Re-Identifying the Temporal Dimension}

This section re-identifies the temporal dimension of \texttt{YJMob100K} by mapping its 75 consecutive days of mobility traces to exact calendar dates. Rather than treating time as anonymous indices, we ground the dataset in real chronology using population-scale behavioral regularities and distinctive temporal anchors.

\begin{figure}[t!]
    \centering
    \includegraphics[width=\linewidth]{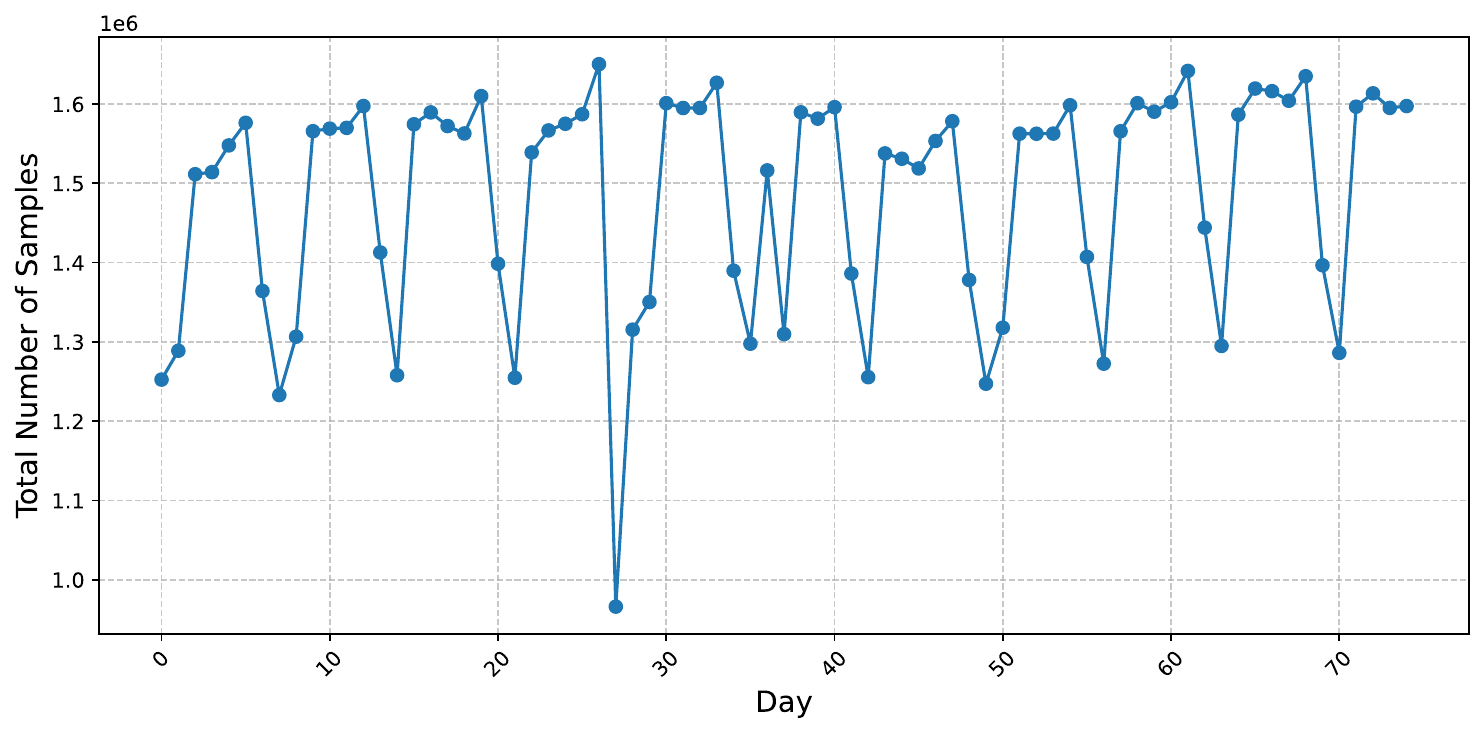}
    \vspace{-0.6cm}
    \caption{Overall activity.}
    \label{fig:general_activity}
\end{figure}

Focusing on Nagoya, the aggregated user activity across all cells (Figure~\ref{fig:general_activity}) exhibits clear weekly periodicity and sharp anomalies. These signals encode unique temporal signatures, including workdays, weekends, public holidays, and major events. By aligning these signatures with publicly known Japanese holidays and disruptions, we infer the dataset's time frame. The dataset publication timeline~\cite{yabe2024YJMob100K} bounds the latest possible end date, which we assume as no later than April 18, 2024, and we bound the earliest possible start date after January 1, 2015. Within this search interval, a single candidate sequence matches both the observed regularity and the exceptional deviations, which enables a precise temporal re-identification.

\begin{table*}[t!]
\centering
\footnotesize
\caption{List of the 75 days in the dataset, along with their class (A -- non-working day / B -- working day) and the inferred weekday. Non-working days outside of weekends, $\mathcal{H}$, are highlighted in yellow.}
\label{tab:days}
\begin{tabular}{|l*{15}{c}|} \hline
\textbf{Day} 
& 0 & \cellcolor{yellow}1 & 2 & 3 & 4 & 5 & 6 & 7 & \cellcolor{yellow}8 & 9 & 10 & 11 & 12 & 13 & 14 \\
\textbf{Class} 
& A & \cellcolor{yellow}A & B & B & B & B & A & A & \cellcolor{yellow}A & B & B & B & B & A & A \\
\textbf{Weekday} 
& Sun & \cellcolor{yellow}Mon & Tue & Wed & Thu & Fri & Sat & Sun & \cellcolor{yellow}Mon & Tue & Wed & Thu & Fri & Sat & Sun \\ \hline

\textbf{Day} 
& 15 & 16 & 17 & 18 & 19 & 20 & 21 & 22 & 23 & 24 & 25 & 26 & 27 & 28 & \cellcolor{yellow}29 \\
\textbf{Class} 
& B & B & B & B & B & A & A & B & B & B & B & B & A & A & \cellcolor{yellow}A \\
\textbf{Weekday} 
& Mon & Tue & Wed & Thu & Fri & Sat & Sun & Mon & Tue & Wed & Thu & Fri & Sat & Sun & \cellcolor{yellow}Mon \\ \hline

\textbf{Day} 
& 30 & 31 & 32 & 33 & 34 & 35 & 36 & \cellcolor{yellow}37 & 38 & 39 & 40 & 41 & 42 & 43 & 44 \\
\textbf{Class} 
& B & B & B & B & A & A & B & \cellcolor{yellow}A & B & B & B & A & A & B & B \\
\textbf{Weekday} 
& Tue & Wed & Thu & Fri & Sat & Sun & Mon & \cellcolor{yellow}Tue & Wed & Thu & Fri & Sat & Sun & Mon & Tue \\ \hline

\textbf{Day} 
& 45 & 46 & 47 & 48 & 49 & \cellcolor{yellow}50 & 51 & 52 & 53 & 54 & 55 & 56 & 57 & 58 & 59 \\
\textbf{Class} 
& B & B & B & A & A & \cellcolor{yellow}A & B & B & B & B & A & A & B & B & A \\
\textbf{Weekday} 
& Wed & Thu & Fri & Sat & Sun & \cellcolor{yellow}Mon & Tue & Wed & Thu & Fri & Sat & Sun & Mon & Tue & Wed \\ \hline

\textbf{Day} 
& 60 & 61 & 62 & 63 & 64 & 65 & 66 & 67 & 68 & 69 & 70 & 71 & 72 & 73 & 74 \\
\textbf{Class} 
& B & B & A & A & B & B & B & B & B & A & A & B & B & B & B \\
\textbf{Weekday} 
& Thu & Fri & Sat & Sun & Mon & Tue & Wed & Thu & Fri & Sat & Sun & Mon & Tue & Wed & Thu \\ \hline
\end{tabular}
\end{table*}

\begin{figure*}[t!]
    \centering
    \includegraphics[width=0.99\linewidth]{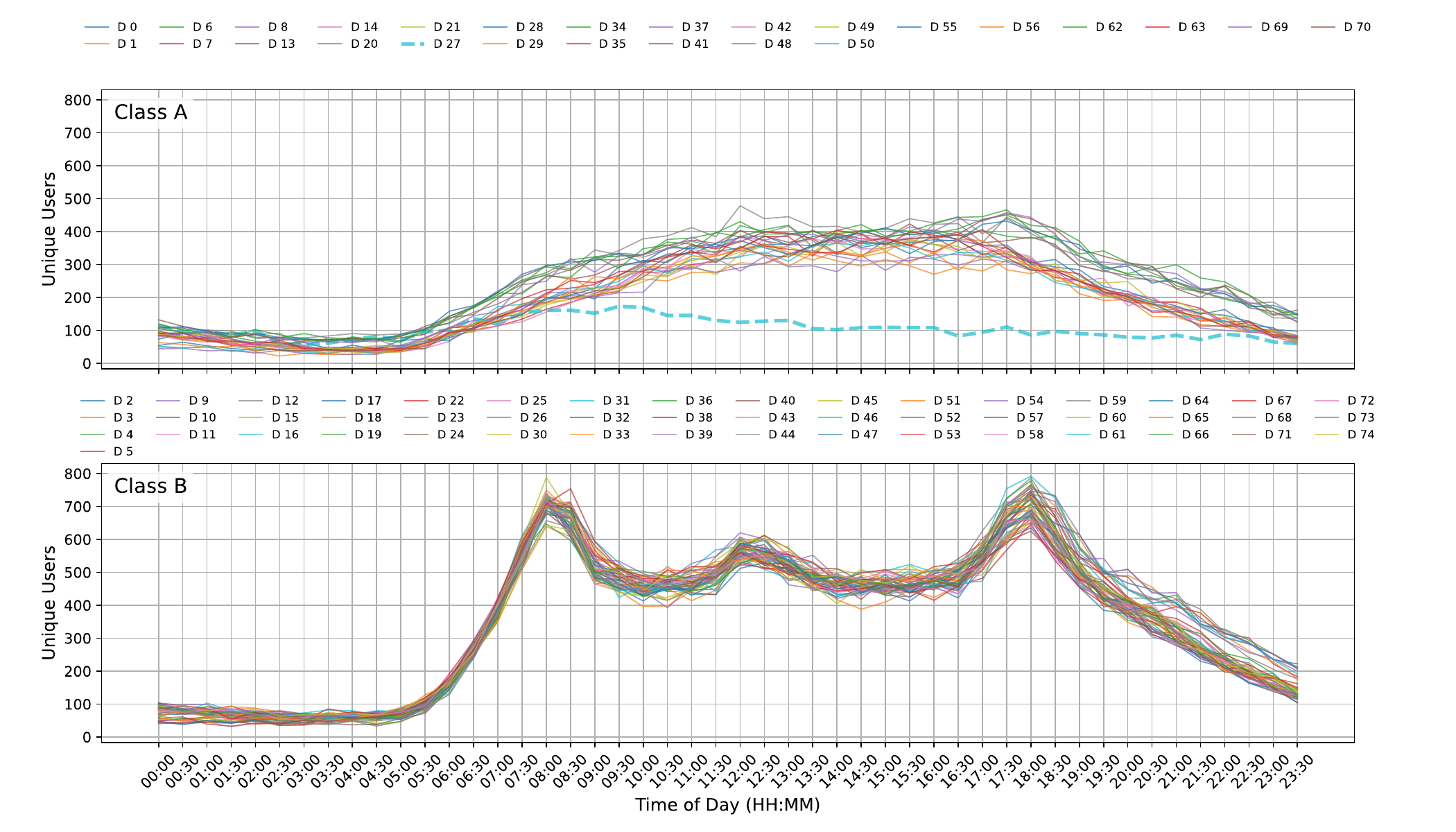}
    \caption{Daily temporal activity in top 10 residential areas organized into two classes from a 2-class clustering.}
    \label{fig:daily_activity_residential_areas}
\end{figure*}

\subsection{Identifying Weekdays}

Figure~\ref{fig:general_activity} shows strong 7-day periodicity in global mobility, which signals a weekly cycle that alternates between working and non-working days. The sequence also contains deviations that indicate public holidays embedded in the data.

To strengthen day-type inference, we analyze activity in the top 10 residential areas. We identify the top grid cells by user count as follows: (82.0, 135.0), (77.0, 135.0), (81.0, 135.0), (82.0, 149.0), (77.0, 134.0), (87.0, 141.0), (80.0, 127.0), (52.0, 185.0), (101.0, 135.0) and (88.0, 124.0). We cluster daily activity profiles using K-Means (with $K=2$ and a random initialization seed), after scaling the 48-bin time series using z-score normalization. Each day corresponds to a 48-dimensional vector of unique user counts per half-hour, and clustering uses Euclidean distance in this normalized space.

Figure~\ref{fig:daily_activity_residential_areas} reveals two distinct behavioral regimes. Class A days show smoother morning ramps and gradual evening declines, consistent with relaxed or irregular routines that typically arise on weekends or holidays. Class B days show sharp morning and evening peaks that reflect commuting work schedules.

Day 27 stands out with unusually low activity and aligns with Class A, which signals a major disruption. The full day labeling (Table~\ref{tab:days}) follows a regular pattern of five consecutive Class B days followed by two Class A days, which confirms a standard Monday to Friday workweek with weekends every seventh day.

Based on this structure, we infer that the dataset begins on a Sunday (Day 0), which enables a deterministic mapping between day indices and weekdays. This weekday alignment enables the public-holiday matching step and supports event-level corroboration later in this section.

\subsection{Identifying the Exact Date}

Building on weekday inference, we now identify exact calendar dates by matching detected non-working days with Japan's public holiday calendar. In addition to regular weekends, we detect five non-working days outside the typical Saturday-Sunday pattern: Day 1 (Monday), Day 8 (Monday), Day 29 (Monday), Day 37 (Tuesday), and Day 50 (Monday). We denote this set of suspected public holidays as $\mathcal{H}$ (highlighted in Table~\ref{tab:days}). 

To pinpoint the dataset timeline, we use Day 37, a Tuesday non-working day, as a strong anchor. We hypothesize that Day 37 aligns with a national holiday and we search for matching sequences in the 2015--2024 range using the official Japanese holiday calendar~\cite{japan_holidays}. For each candidate Tuesday holiday, we test whether the remaining dates in $\mathcal{H}$ match known holidays under the weekday alignment implied by Table~\ref{tab:days}.

A single sequence satisfies all constraints: the dataset begins on 15 September 2019 (Day 0). Under this alignment, $\mathcal{H}$ matches Japan’s national holidays exactly: \emph{Respect for the Aged Day} (16/09/2019 - Day 1), \emph{Autumn Equinox} (23/09/2019 - Day 8), \emph{Health and Sports Day} (14/10/2019 - Day 29), \emph{Enthronement Ceremony Day} (22/10/2019 - Day 37), and \emph{Culture Day Observed} (04/11/2019 - Day 50). This unique match fixes Day 0 as 15/09/2019 and assigns calendar dates to the full 75-day sequence.

\subsection{Additional Timeline Validation}

We further validate the temporal reconstruction by matching real-world events in Nagoya to distinctive patterns in the dataset. These anchors, including disruptions and large gatherings, provide converging evidence that Day 0 corresponds to 15 September 2019.

\subsubsection{Activity anomaly on Day 27: Typhoon Hagibis}

Day 27 (Saturday) exhibits a pronounced activity collapse in Figure~\ref{fig:general_activity}. The top 10 residential areas show an even sharper drop: Day 27 stands as the lowest-activity day in the 75-day sequence (Figure~\ref{fig:daily_activity_residential_areas}).

Under our alignment, Day 27 corresponds to 12 October 2019, the day Typhoon Hagibis makes landfall in Japan. The storm triggers nationwide disruption across transportation and public services, including impacts in the Nagoya region~\cite{bbc_typhoon_hagibis}. This tight temporal correspondence between a major disruption and a dataset-wide behavioral anomaly strongly corroborates our reconstructed timeline.

\begin{figure}[t!]
    \centering
    \includegraphics[width=\linewidth]{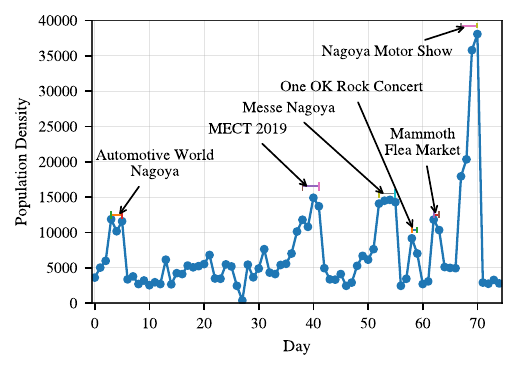}
    \centering
    \vspace{-0.45cm}
    \caption{Activity at Port Messe Nagoya (grid 69,110; GPS 35.05016252379402, 137.39573944062153) with hosted events.}
    \label{fig:port_messe_69_110}
\end{figure}

\subsubsection{Event correlation at Port Messe Nagoya}

Port Messe Nagoya, one of the city's largest exhibition venues, provides an additional temporal anchor. Activity at this location (Figure~\ref{fig:port_messe_69_110}) shows multiple peaks during the dataset period. Each peak aligns with dates of major public events hosted at the venue. By cross-referencing public event calendars, we identify six events at Port Messe during 15/09/2019 to 28/11/2019, and each event coincides with a surge in user presence (see Table~\ref{tab:events_PortMesse} in Section \ref{events_table}).

Furthermore, the magnitude of each peak correlates with the event’s expected attendance. For example, the largest peak aligns with the \emph{Nagoya Motor Show}, which attracts approximately 205{,}900 visitors. This relationship connects the dataset’s spatio-temporal signals to real footfall at a known venue.

Taken together, these anchors, including public holidays, a large-scale disaster disruption, and location-specific event spikes, provide compelling corroboration. We conclude that the dataset begins on 15 September 2019, and our approach re-identifies the exact dates embedded in the anonymized trajectories.

\section{Privacy Risks for Users in \texttt{YJMob100K}}
\label{sec:privacy_risks}

After establishing that we can reverse the protection strategy for \texttt{YJMob100K}, we quantify the resulting privacy risks under the recovered location and timeline. To ensure ethical compliance, we refrain from deeper individualized tracing or identity linkage, even though the reconstructed spatio-temporal frame enables such steps. Prior work~\cite{montjoye_unique_2013} highlights how these analyses risk serious harm when they connect mobility traces to identifiable individuals. 

We introduce state-of-the-art privacy risk metrics that directly quantify leakage once an adversary recovers real coordinates and dates. We design each metric to (i) reflect a concrete adversary capability, (ii) admit rigorous mathematical definition, and (iii) support reproducible measurement and visualization.

\subsection{Threat Model and Notation}
\label{subsec:privacy_threat_model}

We consider an adversary who obtains the released \dataset{} trajectories and applies our spatial and temporal re-identification pipeline. The adversary therefore operates in the real-world reference frame: each record maps to a real grid cell and a real timestamp. The adversary also holds auxiliary knowledge about a target user, such as a home neighborhood, a work area, a small set of spatio-temporal check-ins, or attendance at a public event. This threat model matches the standard linkage setting for mobility privacy~\cite{montjoye_unique_2013,douriez_anonymizing_2016}.

\paragraph{Discrete representation.}
We represent the mobility trace of user $u$ as a sequence
\begin{equation}
\mathcal{T}_u = \{(\ell_{u,i}, \tau_{u,i})\}_{i=1}^{n_u},
\end{equation}
where $\ell_{u,i} \in \mathcal{L}$ denotes a spatial cell (after re-identification) and $\tau_{u,i} \in \mathcal{H}$ denotes a time bin. We use 48 half-hour bins per day when we study daily temporal profiles, and we use day indices $d \in \{0,\dots,74\}$ when we study calendar-level signals.

\paragraph{Visits and unique visitors.}
For any cell $\ell \in \mathcal{L}$, let
\begin{equation}
\mathcal{U}(\ell) = \{u \mid \exists i \text{ s.t. } \ell_{u,i}=\ell \}
\quad \text{and} \quad
s(\ell)=|\mathcal{U}(\ell)|
\end{equation}
denote the set and the count of unique users who have ever visited $\ell$. We use $s(\ell)$ as a principled measure of spatial seclusion.

\subsection{Spatio-Temporal k-Anonymity}
\label{subsec:k_anonymity}

A dataset leaks identity when small fragments of a trace isolate a single user. We capture this effect through a spatio-temporal k-anonymity metric that counts how many users remain plausible matches under an adversary query.

\paragraph{Adversary query model.}
Let a query $Q$ consist of $m$ spatio-temporal constraints:
\begin{equation}
Q = \{(\ell_j, \Delta \tau_j)\}_{j=1}^{m},
\end{equation}
where each constraint says: ``the user appears in spatial cell $\ell_j$ during time window $\Delta \tau_j$.'' Time windows let us model realistic uncertainty (for example, ``around 9am'' or ``during lunch'').

We define the \emph{candidate set} of users consistent with $Q$ as
\begin{equation}
\mathsf{Cand}(Q) = \left\{u \,\middle|\,
\forall (\ell_j,\Delta \tau_j)\in Q,\ \exists i:\ \ell_{u,i}=\ell_j \wedge \tau_{u,i} \in \Delta \tau_j
\right\}.
\end{equation}
We define the \emph{k-anonymity of query $Q$} as
\begin{equation}
k(Q) = |\mathsf{Cand}(Q)|.
\end{equation}

\paragraph{Population risk summary.}
We report the fraction of queries that yield small candidate sets:
\begin{equation}
\mathsf{Risk}_k(m,\Delta) = \Pr_{Q \sim \mathcal{D}(m,\Delta)}\left[ k(Q) \le k \right],
\end{equation}
where $\mathcal{D}(m,\Delta)$ samples $m$ constraints from real user traces and uses a time-window size $\Delta$ (for example, $\Delta=30$ minutes or $\Delta=2$ hours). This metric answers a critical security question: \emph{how many auxiliary check-ins does an attacker need to isolate a user?}

\begin{figure}[t!]
  \centering
  \subfigure{
    \includegraphics[width=0.98\linewidth]{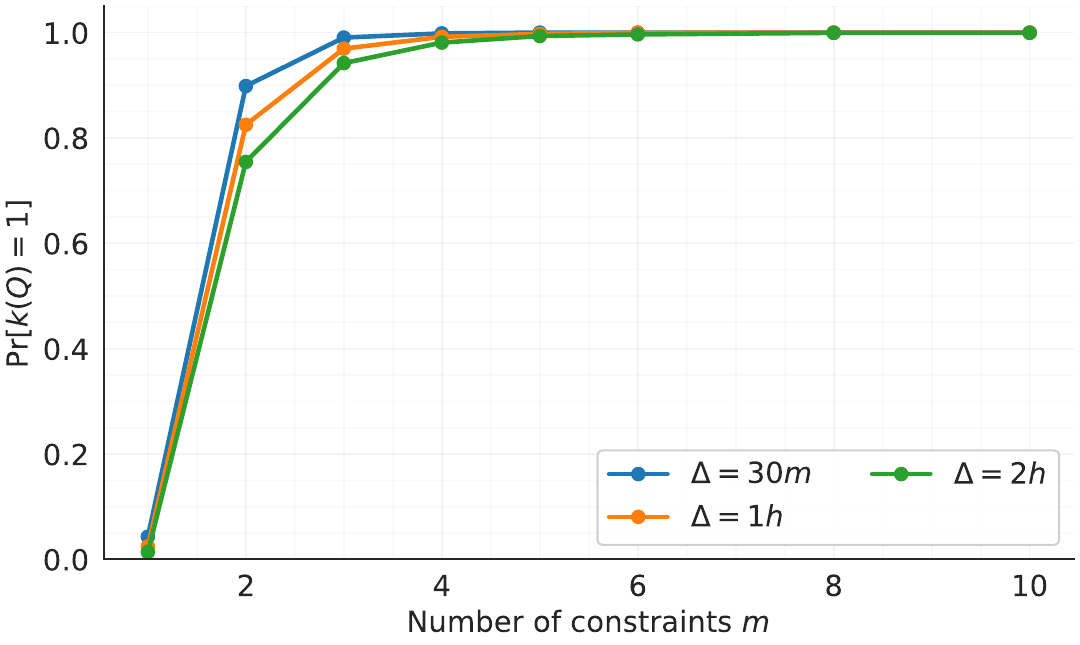}
    \label{fig:k_uniqueness_vs_m}
  }
  \vspace{-0.45cm}
  \caption{Spatio-temporal $k$-anonymity risk under realistic adversary queries. Uniqueness probability rapidly increases with the number of spatio-temporal constraints $m$, even with coarse temporal tolerance $\Delta$. }
  \label{fig:metric_k_anonymity}
\end{figure}

Figure~\ref{fig:metric_k_anonymity} shows that only a few spatio-temporal constraints are enough to isolate users even with coarse temporal tolerance $\Delta$. In Fig.~\ref{fig:k_uniqueness_vs_m}, the uniqueness probability $\Pr[k(Q)=1]$ is small for $m{=}1$ (only a few percent, depending on $\Delta$), but it jumps dramatically at $m{=}2$ (already around $0.75$--$0.90$ as $\Delta$ tightens) and exceeds $\approx 0.94$ by $m{=}3$ for all shown $\Delta$, after which it rapidly saturates close to 1. 

\subsection{Trajectory Unicity Under Partial Knowledge}
\label{subsec:unicity}

The mobility privacy literature repeatedly shows that a small number of points uniquely identifies individuals~\cite{montjoye_unique_2013}. After re-identification, \dataset{} supports the same style of attack at the city scale and with real dates. We therefore quantify \emph{unicity} directly, and we do so in a way that matches our discrete representation.

\paragraph{Point-set representation.}
Let $\mathsf{Set}(\mathcal{T}_u)$ denote the set of distinct spatio-temporal points visited by $u$:
\begin{equation}
\mathsf{Set}(\mathcal{T}_u)=\{(\ell,\tau)\mid \exists i:\ (\ell_{u,i},\tau_{u,i})=(\ell,\tau)\}.
\end{equation}

\paragraph{m-point unicity of a user.}
For a fixed $m$, sample a subset $P_u^{(m)} \subset \mathsf{Set}(\mathcal{T}_u)$ with $|P_u^{(m)}|=m$. We define the match count
\begin{equation}
M(P_u^{(m)}) = \left|\left\{v \mid P_u^{(m)} \subseteq \mathsf{Set}(\mathcal{T}_v)\right\}\right|.
\end{equation}
We call $u$ \emph{unique under $m$ points} when $M(P_u^{(m)})=1$.

\paragraph{Population unicity curve.}
We define the dataset-level unicity as
\begin{equation}
U(m) = \Pr_{u \sim \mathcal{U},\ P_u^{(m)}}\left[M(P_u^{(m)})=1\right],
\end{equation}
where $\mathcal{U}$ denotes the uniform distribution over users and $P_u^{(m)}$ denotes uniform sampling over $m$-subsets from $\mathsf{Set}(\mathcal{T}_u)$.

$U(m)$ quantifies the re-identification surface directly: a high $U(m)$ means an attacker who learns $m$ check-ins (for example, from posts, receipts, or observed commutes) isolates a user with high probability.

\begin{figure}[t!]
  \centering
  \includegraphics[width=0.95\linewidth]{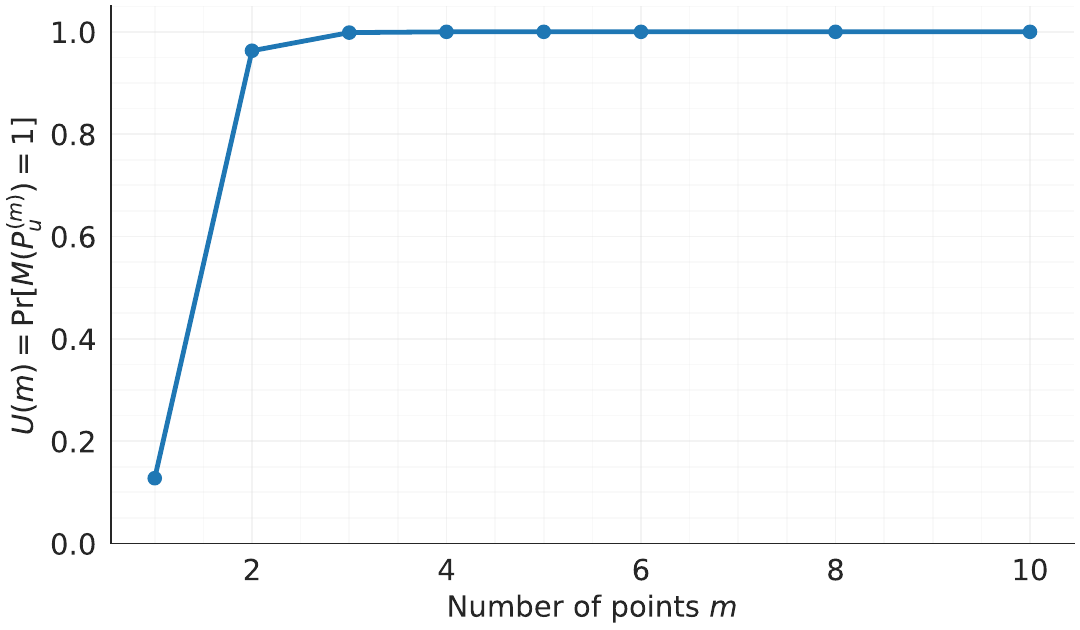}
  \vspace{-0.45cm}
  \caption{Trajectory unicity under as a function of the number of known spatio-temporal points $m$.}
  \label{fig:metric_unicity}
\end{figure}

\begin{figure*}[t!]
  \centering
  \subfigure{
    \includegraphics[width=0.32\linewidth]{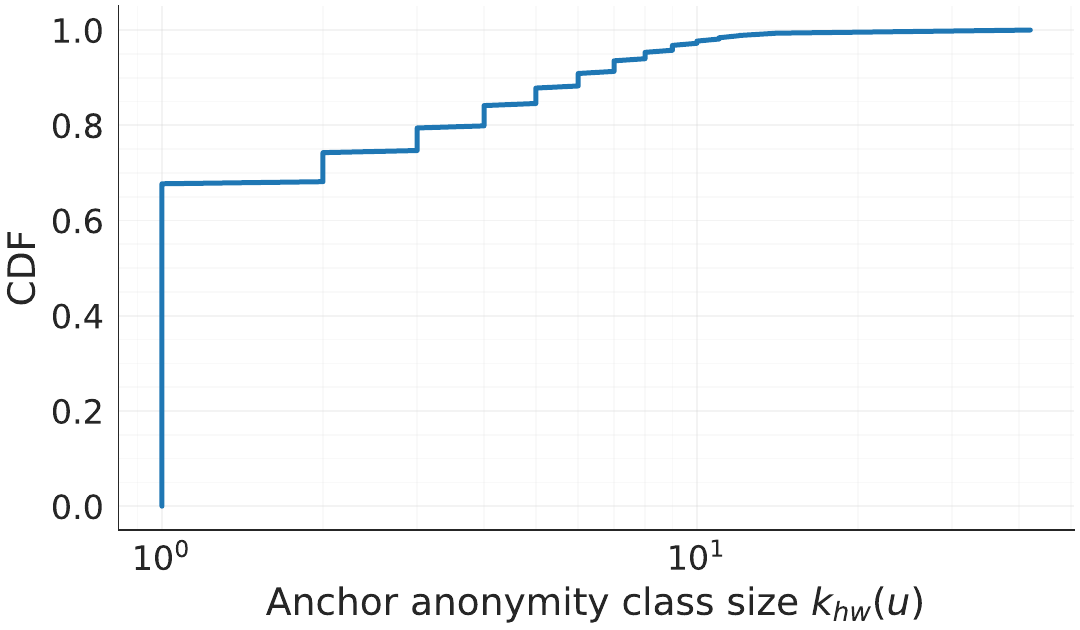}
    \label{fig:k_hw_cdf}
  }
  \hfill
  \subfigure{
    \includegraphics[width=0.32\linewidth]{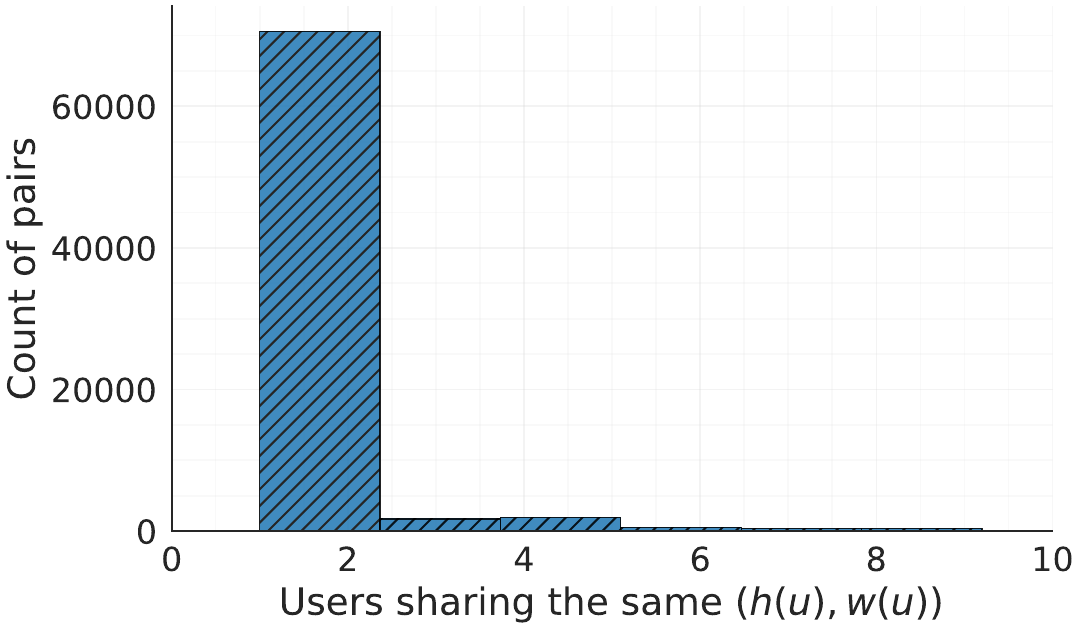}
    \label{fig:hist_shared_hw}
  }
  \hfill
  \subfigure{
    \includegraphics[width=0.32\linewidth]{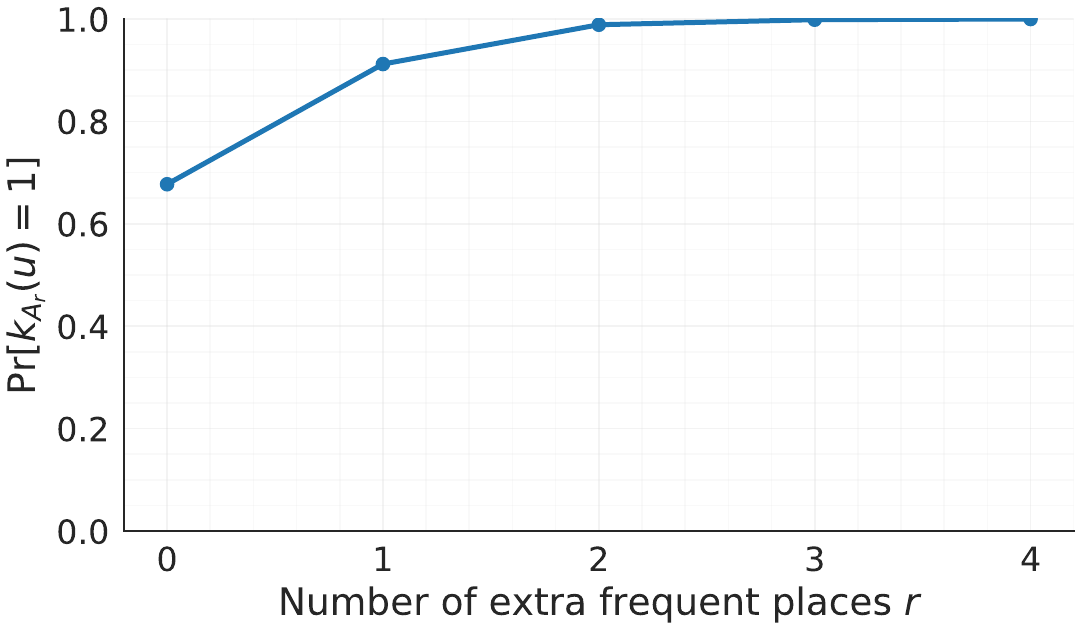}
    \label{fig:uniq_vs_r}
  }
  \vspace{-0.45cm}
  \caption{Anchor-based privacy leakage. (a) Distribution of home--work anonymity class sizes $k_{hw}(u)$, where smaller values indicate users whose home and work pair is shared with few others. (b) Histogram of the number of users sharing each $(h(u),w(u))$ pair, highlighting how many home--work combinations are effectively unique or very sparsely shared. (c) Probability that the full anchor signature $\mathbf{A}_r(u)=(h(u),w(u),p_1(u),\dots,p_r(u))$ uniquely identifies a user as the attacker learns additional frequent non-home, non-work places.}
  \label{fig:metric_anchor_uniqueness}
\end{figure*}

Figure~\ref{fig:metric_unicity} reports the unicity curve, i.e., the probability that a randomly chosen user $u$ becomes uniquely identifiable when an adversary knows $m$ sampled spatio-temporal points $P_u^{(m)}\subset\mathsf{Set}(\mathcal{T}_u)$. The curve rises sharply: with only one point, unicity is low (about $0.13$), but with two points it jumps to near-certain re-identification (about $0.97$), and it essentially saturates to $U(m)\approx 1$ from $m\ge 3$. This indicates that, in this dataset and discretization, very small partial knowledge (only a couple of check-ins) is sufficient to isolate almost any individual trajectory, confirming the strong linkage surface captured by $M(P_u^{(m)})$ and the definition of $U(m)$. 

\subsection{Anchor-based Uniqueness}
\label{subsec:anchors}

Home and work anchors drive linkage attacks because they provide stable, repeated signals. After re-identification, these anchors correspond to real neighborhoods and real routines. We therefore define metrics that quantify how strongly anchors isolate users.

\paragraph{Home and work operators.}
Let $\mathcal{T}_u$ contain visits across days. Define two time masks:
\begin{equation}
\mathcal{H}_\text{home}=\{\tau \mid \tau \in [22{:}00,06{:}00)\} 
\end{equation}
\begin{equation}
\mathcal{H}_\text{work}=\{\tau \mid \tau \in [09{:}00,17{:}00)\}
\end{equation}
Define the home and work cells as
\begin{equation}
h(u)=\arg\max_{\ell \in \mathcal{L}} \#\{i \mid \ell_{u,i}=\ell \wedge \tau_{u,i}\in \mathcal{H}_\text{home}\}
\end{equation}
\begin{equation}
w(u)=\arg\max_{\ell \in \mathcal{L}} \#\{i \mid \ell_{u,i}=\ell \wedge \tau_{u,i}\in \mathcal{H}_\text{work}\}
\end{equation}

\paragraph{Home-work uniqueness.}
We define the equivalence class of a user under anchors as
\begin{equation}
\mathcal{E}_{hw}(u)=\{v \mid h(v)=h(u) \wedge w(v)=w(u)\},
\quad
k_{hw}(u)=|\mathcal{E}_{hw}(u)|.
\end{equation}
We report the \emph{unique-anchor rate}
\begin{equation}
\mathsf{UA}_{hw} = \frac{1}{|\mathcal{U}|}\sum_{u\in\mathcal{U}} \mathbf{1}[k_{hw}(u)=1],
\end{equation}
and we also report the distribution of $k_{hw}(u)$ to show how quickly anonymity collapses as an attacker learns anchors.

\paragraph{Multi-anchor extension.}
Real attackers often know more than home and work. Let $p_1(u),\dots,p_r(u)$ denote the top-$r$ most visited non-home, non-work cells for user $u$ over the full horizon. We define the anchor signature
\begin{equation}
\mathbf{A}_r(u) = \big(h(u), w(u), p_1(u),\dots,p_r(u)\big).
\end{equation}
We define
\begin{equation}
k_{A_r}(u)=\left|\left\{v \mid \mathbf{A}_r(v)=\mathbf{A}_r(u)\right\}\right|.
\end{equation}
We report $\Pr[k_{A_r}(u)=1]$ as a function of $r$. This metric captures a realistic escalation: each additional frequent place acts as an extra key in a linkage attack. 

Figure~\ref{fig:metric_anchor_uniqueness} shows that home and work already act as strong quasi-identifiers, and that a few extra anchors almost completely collapse anonymity. From panel~\ref{fig:k_hw_cdf}, we find that about $\approx 68\%$ of users have $k_{hw}(u)=1$ (unique home--work pair), and roughly $\approx 88\%$ satisfy $k_{hw}(u)\le 5$, so most users share their $(h(u),w(u))$ combination with at most a handful of others. Panel~\ref{fig:hist_shared_hw} confirms that many anchor pairs are highly specific: around $\approx 68\%$ of distinct $(h(u),w(u))$ pairs are used by a single user, and over $\approx 87\%$ by at most three users. The multi-anchor curve in panel~\ref{fig:uniq_vs_r} then shows how this risk escalates as we include more frequent locations: the uniqueness probability jumps from $\Pr[k_{A_0}(u)=1]\approx 0.68$ (home+work only) to $\approx 0.91$ with one extra place ($r=1$), exceeds $\approx 0.98$ for $r=2$, and reaches essentially full uniqueness ($>0.99$) by $r=3$–$4$, indicating that knowing just a few stable locations beyond home and work is enough to uniquely pinpoint almost every user. 

\subsection{Seclusion Exposure}
\label{subsec:seclusion_exposure}

Seclusion creates high-risk moments: sparse cells allow easy isolation, and visits to sparse cells often correspond to sensitive places and increase individual re-identification of users. We therefore quantify \emph{how much} of each trajectory occurs in low-support regions.

\paragraph{Per-user seclusion exposure.}
For a threshold $\kappa$ (for example, 1, 3, 10 users per cell), we define the per-user seclusion exposure ratio:
\begin{equation}
\mathsf{SE}_\kappa(u)=
\frac{\sum_{i=1}^{n_u} \mathbf{1}\big[s(\ell_{u,i}) \le \kappa\big]}{n_u}.
\end{equation}
This metric measures the fraction of a user’s samples that fall in cells with at most $\kappa$ unique visitors.

\begin{figure}[t!]
\centering
\subfigure{
\includegraphics[width=0.98\linewidth]{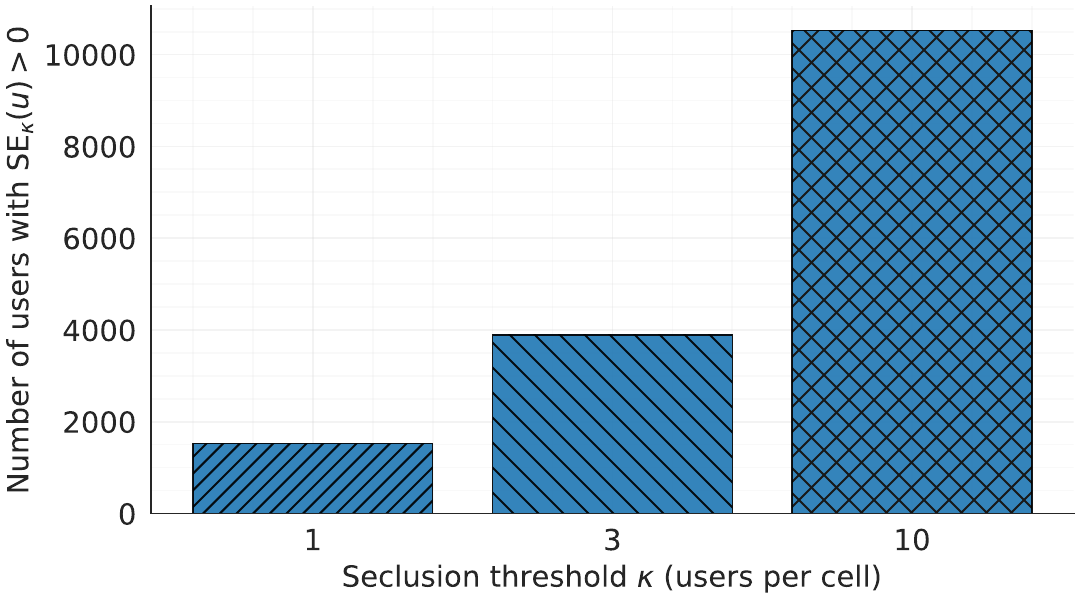}
\label{fig:secl_exposed_bar}
}
\vspace{-0.45cm}
\caption{Users in cells with seclusion thresholds $\kappa$.}
\label{fig:metric_seclusion}
\end{figure}

Figure~\ref{fig:metric_seclusion} summarizes how much of each trajectory runs through low-support regions. It shows that a considerable number of users visit truly isolated cells: about 2,000 users have $\mathsf{SE}*1(u)>0$, rising to roughly 4,000 for $\kappa=3$ and around 11,000 for $\kappa=10$, signifying a high individual re-identification threat. 

\subsection{Sensitive-POI Disclosure}
\label{subsec:sensitive_poi}

After re-identification, a cell maps to a physical region. This mapping enables semantic inference because public POI catalogs label categories such as hospitals, places of worship, nightlife, political venues, and support groups (listed in Section \ref{sensitive_pois}). We quantify risk without performing individual identification by measuring exposure and uniqueness of sensitive visits.

\paragraph{Sensitive categories.}
Let $\mathcal{S} \subseteq \mathcal{L}$ denote the set of cells that intersect sensitive PoIs under a public taxonomy (for example, medical, religious, sexual health, addiction support, or political organizations). Let $\mathcal{C}$ denote categories, and let $c(\ell)\in \mathcal{C}$ denote the category label of a cell when available.

\paragraph{Sensitive signature uniqueness.}
Attackers often use a small set of rare venues as quasi-identifiers. We define the top-$q$ sensitive cells of user $u$:
\begin{equation}
\mathcal{S}_q(u)=\text{Top-}q \text{ cells in } \{\ell \in \mathcal{S} \mid \ell \text{ appears in } \mathcal{T}_u\} \text{ by visit count.}
\end{equation}
We define the sensitive uniqueness class size:
\begin{equation}
k_{\text{sens}}(u)=\left|\left\{v \mid \mathcal{S}_q(v)=\mathcal{S}_q(u)\right\}\right|.
\end{equation}
We report $\Pr[k_{\text{sens}}(u)=1]$ for small $q$ (for example, $q\in\{1,2,3\}$). This metric captures a concrete leakage channel: a small set of sensitive visits uniquely pins down a user even without home and work.

\begin{figure}[t]
\centering
\subfigure{
\includegraphics[width=0.98\linewidth]{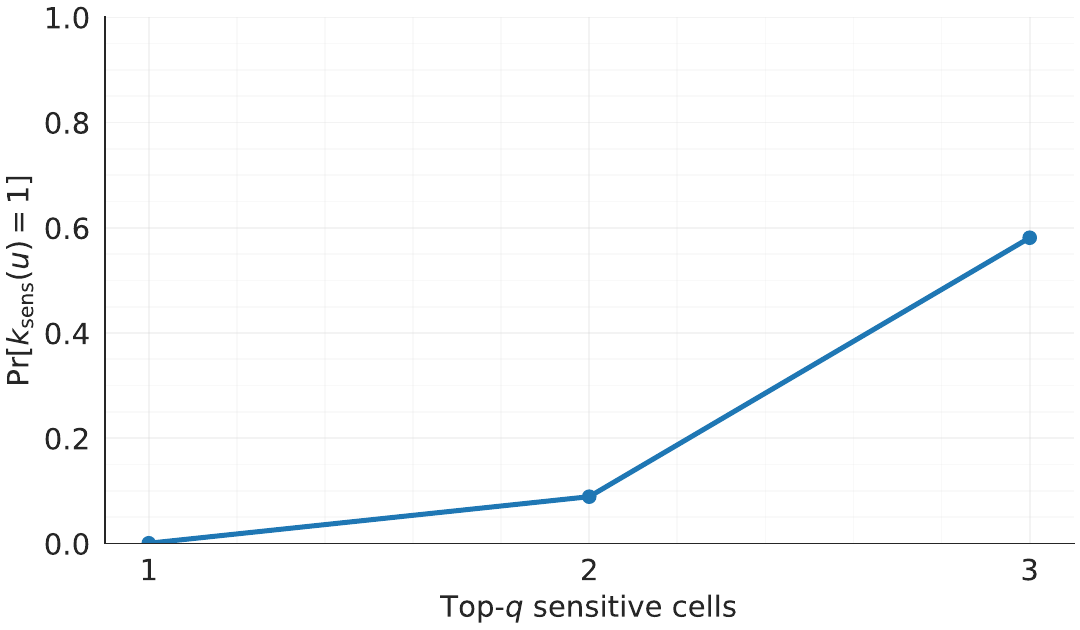}
\label{fig:metric_sensitive_poi_uniqueness_vs_q}
}
\vspace{-0.45cm}
\caption{Sensitive-POI uniqueness. Probability that the top-$q$ sensitive cells $\mathcal{S}_q(u)$ uniquely identify a user.}
\label{fig:metric_sensitive_poi}
\end{figure}

Figure~\ref{fig:metric_sensitive_poi} quantifies how often users appear near sensitive venues and how quickly such visits become identifying. It focuses on users who ever visit such cells and tracks the sensitive signature uniqueness $\Pr[k_{\text{sens}}(u)=1]$ as we reveal the top-$q$ sensitive cells $\mathcal{S}_q(u)$. A single sensitive cell is almost never uniquely identifying (uniqueness well below $1\%$), but adding one more already makes around $9\%$ of these users unique, and by $q=3$, more than half of them are uniquely pinned down by their top three sensitive locations alone. Thus, a small set of frequently visited sensitive cells quickly acts as a powerful quasi-identifier.

\section{Protecting Mobility Datasets like \texttt{YJMob100K}} 

Having established that the protection measures applied to the dataset are ineffective, we investigate if other sanitization schemes can provide satisfactory protection, specifically in the spatial dimension. We start by applying state-of-the-art schemes that are working at the local level by modifying the individual traces. Then, we consider a more global and destructive approach by removing the spatial structure. 

\subsection{Possible Sanitization Techniques}

We evaluate three representative privacy-preserving mechanisms that span different points in the design space for trajectory anonymization: \geoind{}~\cite{Chatzikokolakis2013} for point-wise differential privacy, Generalized Randomized Response (GRR)~\cite{kairouz2016discrete} for location-level local differential privacy, and spatial de-structuring to eliminate global spatial coherence. We select these techniques for their theoretical foundations, empirical relevance, and practical scalability to large-scale mobility datasets. \geoind{} targets quality-preserving perturbation for local services. GRR targets aggregate analyses such as flow estimation under local privacy. Spatial de-structuring provides an aggressive baseline: it preserves density profiles while discarding spatial continuity.

Recent methods such as PrivTrace~\cite{wang2023privtrace} aim for trajectory-level privacy using Markov modeling and convex optimization. However, these approaches often impose computational complexity that challenges large-scale reproducibility. In particular, the overall time complexity scales as $\mathcal{O}(m|D| + m^3)$, where $m$ denotes the number of spatial cells and $|D|$ denotes the number of trajectories. This complexity arises from discretization, shortest-path estimation, and higher-order transition modeling, and it becomes difficult to scale to datasets like \texttt{YJMob100K} with high spatial granularity and very large user counts. In contrast, our selected techniques balance theoretical footing, interpretability, and feasibility for large-scale evaluation. We also note that synthetic trajectory generation constitutes a complementary baseline. Our focus here is on stress-testing commonly deployed on-device sanitization mechanisms, while the proposed framework also extends naturally to evaluating synthetic approaches against structure-based re-identification. 

\subsection{Geo-Indistinguishability}

In this study, we employed Geo-indistinguishability (\geoind{})~\cite{Andrs2013} to anonymize individual location points within user trajectories. \geoind{} offers strong theoretical guarantees and is widely regarded as the \emph{de facto} standard for location privacy in the absence of a trusted data curator. Its design extends Differential Privacy (DP)~\cite{Dwork2006Calibrating} to the spatial domain, ensuring that nearby locations yield statistically indistinguishable outputs, thereby limiting the adversary’s ability to infer precise user positions.

Our choice to use \geoind{} is motivated by its established role in the literature and by recent evidence that re-identification risks remain high in the absence of formal privacy guarantees. However, while \geoind{} effectively obfuscates individual locations, protecting complete mobility trajectories remains an open challenge. As emphasized in the recent survey by Miranda-Pascual et al.~\cite{MirandaPascual2023}, achieving DP for trajectory data is particularly difficult due to the sequential and highly correlated nature of spatiotemporal movements. The authors highlight that existing techniques often struggle to simultaneously ensure privacy and maintain utility and that designing privacy-preserving mechanisms capable of protecting full trajectories is still an unresolved research problem. 

More formally, a mechanism satisfies $\epsilon$-\geoind{} if for any two locations $x_1,x_2\in \mathbb{R}^2$ within a radius $r$, and any possible output $y \in \mathbb{R}^2$, we have:
\begin{equation*}\label{eq:geo}
\frac{\Pr(y|x_1)}{\Pr(y|x_2)} \leq e^{\epsilon r} \textrm{, } \forall r>0 \textrm{, } \forall y \textrm{, } \forall x_1,x_2: d(x_1,x_2)\leq r\textrm{.}
\end{equation*}

Intuitively, for any point $x_2$ within distance $r$ of $x_1$, \geoind{} ensures that their output distributions are close-bounded by $l=\epsilon r$. For example, an adversary might distinguish that a user is in Paris rather than London, but may struggle to identify a specific street in Paris, depending on $\epsilon$. Although \geoind{} and classical DP use $\epsilon$ to denote a privacy budget, they are not directly comparable: in \geoind{}, $\epsilon$ is scale-dependent and incorporates spatial units (e.g., meters$^{-1}$).

In the continuous plane setting considered in this paper, \geoind{} can be instantiated using the planar Laplace mechanism~\cite{Andrs2013}. Rather than reporting the true location $x \in \mathbb{R}^2$, the mechanism samples a perturbed location $y \in \mathbb{R}^2$ using the probability density function:
\[
D_{\epsilon}(y) = \frac{\epsilon^2}{2\pi} e^{-\epsilon d(x, y)} \mathrm{.}
\]
where $d(x, y)$ is the Euclidean distance. The algorithmic implementation of this mechanism is shown in Algorithm~\ref{alg:GI_location}. 

\begin{algorithm}[H]
\caption{Polar Laplace mechanism in continuous plane~\cite{Andrs2013}}
\label{alg:GI_location}
\begin{algorithmic}[1]
\State \textbf{Input :} $\epsilon>0$, real location $x \in \mathbb{R}^2$.
\State \textbf{Output :}  sanitized location $y \in \mathbb{R}^2 $.

\State Draw $\theta$ uniformly in $[0,2\pi)$
\State Draw $p$ uniformly in $[0,1)$
\State Set $r = C^{-1}_{\epsilon}(p) = -\frac{1}{\epsilon} \left ( W_{-1} \left ( \frac{p-1}{e}\right) + 1 \right )$

\State \textbf{Return :} $y=x+\langle r \cos{(\theta)},r \sin{(\theta)} \rangle$
\end{algorithmic}
\end{algorithm} 

\begin{figure}[t!]
    \centering
    \subfigure[Clustered correlation]{\includegraphics[width=0.98\linewidth]{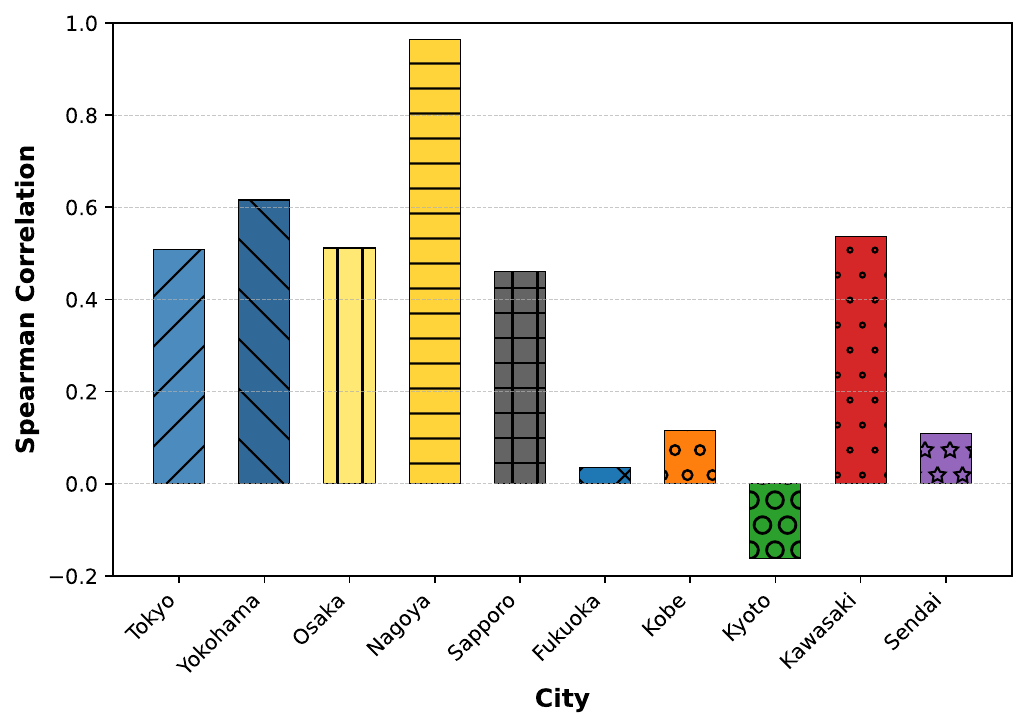} \label{fig:clustered_correlation_noisy_geoind}}
    \hfill
    \vspace{-0.45cm}
    \subfigure[Error in home inference]{\includegraphics[width=0.98\linewidth]{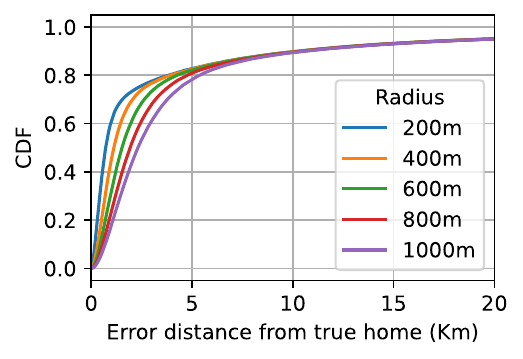} \label{fig:error_home_geoind}}

    \vspace{-0.45cm}
    \caption{(a) Clustered correlation after applying \geoind{} with a 1 km radius; Nagoya remains the dominant match, which supports location re-identification. (b) Error in home inference when comparing pre- and post-perturbation anchors under \geoind{}.}
    \label{fig:correlation_comparison_geoind}
\end{figure}

\begin{figure}[t!]
    \centering
    \subfigure[Re-identification Rate]{%
        \includegraphics[width=0.98 \linewidth]{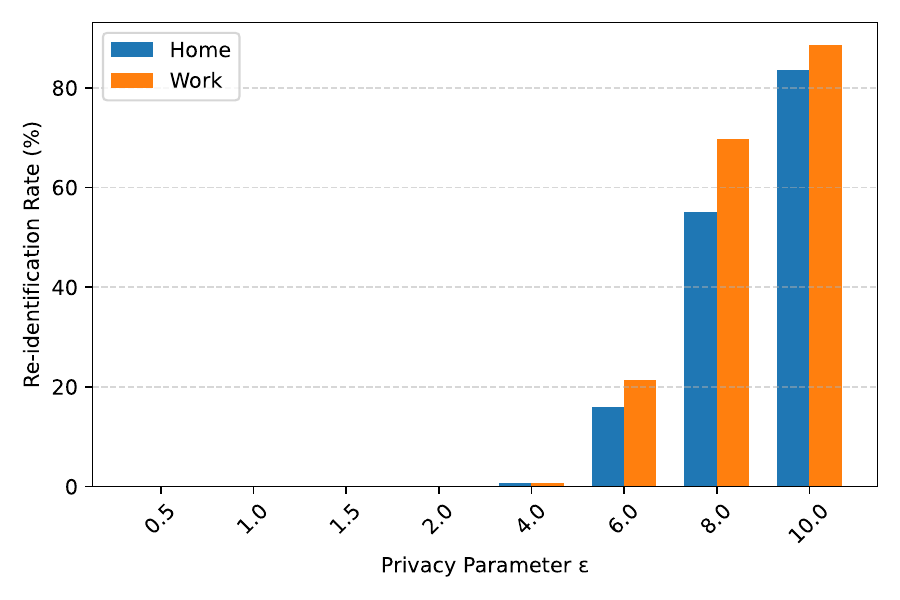}
        \label{fig:reid_vs_epsilon}
    }
    \hfill
    \vspace{-0.45cm}
    \subfigure[KL Divergence]{%
        \includegraphics[width=0.98\linewidth]{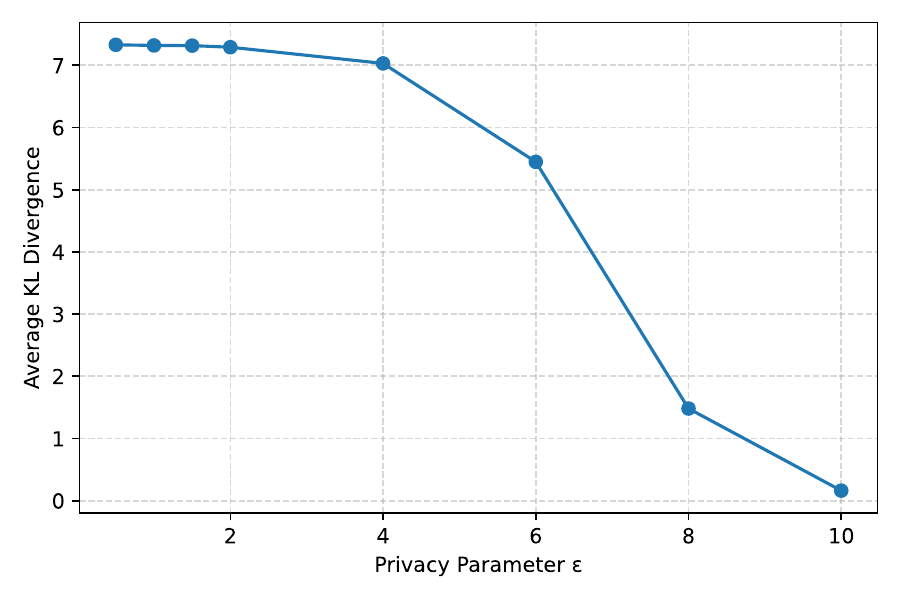}
        \label{fig:kl_vs_epsilon}
    }
    \vspace{-0.45cm}
    \caption{Effect of GRR-based LDP on home–work inference. (a) Home–work re-identification rate vs.\ $\epsilon$. (b) KL divergence between true and perturbed population distributions.}
    \label{fig:ldp_home_work_tradeoff}
\end{figure}

\vspace{0.2cm}
\noindent\textbf{Observations:} We applied \geoind{} at the level of individual location points to assess whether such point-wise perturbations could meaningfully mitigate re-identification risks in the \texttt{YJMob100K} large-scale mobility dataset. To instantiate the mechanism, we used a privacy parameter $\epsilon$ corresponding to a radius of 1 km, which is considered a reasonably strong privacy guarantee in the context of location-based services~\cite{Andrs2013}. This setting introduces a significant amount of spatial uncertainty, yet, as our experiments reveal, it does not fully obscure high-level structural information. 

Figure~\ref{fig:correlation_comparison_geoind}(a) presents the results of our clustered correlation analysis after applying \geoind{} noise. Despite the perturbations, the transformed dataset remains highly correlated with the population distribution and still reveals Nagoya as the recorded city, suggesting that large-scale patterns in user mobility persist and can still be exploited by adversaries. 

In Figure~\ref{fig:correlation_comparison_geoind}(b), we further examine the effectiveness of \geoind{} in obfuscating sensitive points such as home locations. We observe a considerable yet limited increment in the error when varying the privacy parameter by increasing the radius. 

This highlights a critical limitation: while \geoind{} ensures privacy at the level of individual points, the aggregation of these points, especially over repeated visits to the same locations, can leak identifying information. Taken together, these results emphasize that point-wise mechanisms, although theoretically robust, may fall short in practical scenarios involving rich trajectory data. 

\subsection{Generalized Randomized Response (GRR)}

To evaluate the privacy--utility trade-off under a canonical local mechanism, we apply the Generalized Randomized Response (GRR)~\cite{kairouz2016discrete} to protect users’ location trajectories. GRR is one of the most widely studied mechanisms in the local differential privacy (LDP) literature: it offers an analytically simple and optimally private channel for categorical data, and it is routinely used as a baseline when comparing LDP mechanisms. In our setting, the city is discretized into spatial grid cells; once each location is mapped to a discrete symbol, GRR becomes a natural choice to stress-test what can be achieved with point-wise LDP on trajectories without relying on a trusted curator.

\paragraph{Discrete representation and notation.}
Let $\mathcal{L}$ denote the set of spatial cells obtained from tessellating \texttt{YJMob100K}, and let $|\mathcal{L}| = k$. We index cells by integers and write
\[
\phi : \mathcal{L} \rightarrow \{1,\dots,k\}, \quad \ell_{u,i} \in \mathcal{L},\quad z_{u,i} = \phi(\ell_{u,i}) \in \{1,\dots,k\},
\]
where $\ell_{u,i}$ is the $i$-th true location of user $u$ and $z_{u,i}$ its discrete index. GRR acts on this discrete domain $\mathcal{X} = \{1,\dots,k\}$: for a true symbol $x \in \mathcal{X}$, the mechanism outputs a sanitized symbol $\tilde{x} \in \mathcal{X}$.

\paragraph{GRR mechanism and privacy guarantee.}
Fix a privacy parameter $\epsilon > 0$. GRR specifies, for each true value $x \in \{1,\dots,k\}$, the distribution of the reported value $\tilde{x}$:
\begin{equation*}
\Pr[\tilde{x}=v \mid x] =
\begin{cases}
p = \dfrac{e^\epsilon}{e^\epsilon + k - 1}, & \text{if } v = x,\\[0.8em]
q = \dfrac{1}{e^\epsilon + k - 1}, & \text{if } v \neq x.
\end{cases}
\end{equation*}
In words, the mechanism preserves the true cell with probability $p$, and with the remaining probability $1-p$ it outputs a value drawn uniformly from the $k-1$ other cells. This channel is $\epsilon$-locally differentially private: for any two inputs $x,x' \in \{1,\dots,k\}$ and any output $v$,
\[
\frac{\Pr[\tilde{x}=v \mid x]}{\Pr[\tilde{x}=v \mid x']} \le e^\epsilon.
\]
Indeed, the largest ratio occurs between \emph{in-channel} and \emph{out-of-channel} probabilities, and we have $p/q = e^\epsilon$ by construction. Thus, each reported cell $\tilde{x}$ hides its true origin among all $k$ possibilities, with indistinguishability controlled by $\epsilon$. 

\paragraph{Application to trajectories.}
We apply GRR independently to every point in each user’s trajectory. For each sample $(\ell_{u,i}, t_{u,i})$, we compute its discrete index $z_{u,i} = \phi(\ell_{u,i})$, run GRR with input $z_{u,i}$, and obtain a sanitized index $\tilde{z}_{u,i}$. Mapping $\tilde{z}_{u,i}$ back through $\phi^{-1}$ yields a perturbed cell $\tilde{\ell}_{u,i}$. This step produces a synthetic trajectory
\[
\widetilde{\mathcal{T}}_u = \{(\tilde{\ell}_{u,i}, t_{u,i})\}_i
\]
for each user $u$, in which every reported location is protected by $\epsilon$-LDP at the point level.

\paragraph{Evaluation methodology.}
We generate GRR-perturbed datasets for a range of privacy budgets $\epsilon$ and assess both privacy and utility. On the privacy side, we re-run our home/work inference pipeline directly on $\widetilde{\mathcal{T}}_u$ and measure the re-identification rate of anchors $(h(u)$ and $w(u))$ as a function of $\epsilon$, yielding the curve in Figure~\ref{fig:reid_vs_epsilon}. On the utility side, we estimate population-level spatial distributions at each time slot. Let $f_\tau \in \Delta^{k-1}$ denote the true frequency vector over cells at time $\tau$ and $\tilde{f}_\tau$ the empirical distribution obtained from GRR outputs. Since GRR is a linear channel with known parameters $(p,q)$, we apply the standard debiasing estimator for GRR~\cite{kairouz2016discrete} to recover an estimate $\widehat{f}_\tau$ of $f_\tau$, and we quantify distortion via the Kullback--Leibler divergence $\mathrm{KL}(f_\tau \,\|\, \widehat{f}_\tau)$. Figure~\ref{fig:kl_vs_epsilon} reports the average KL divergence across time.

\vspace{0.2cm}
\noindent\textbf{Observations:} Figure~\ref{fig:ldp_home_work_tradeoff} summarizes the trade-off induced by GRR. For smaller privacy budgets (e.g., $\epsilon \le 2$), the home--work re-identification rate in Figure~\ref{fig:reid_vs_epsilon} decreases substantially, indicating that anchor linkage becomes difficult for the attacker; however, Figure~\ref{fig:kl_vs_epsilon} shows that the average KL divergence between true and reconstructed population distributions increases sharply, signaling severe loss of utility for aggregate analyses. As $\epsilon$ grows (e.g., $\epsilon \ge 6$), GRR preserves most of the population structure (low KL divergence), but the re-identification rate climbs back toward the original level, meaning that anchor privacy is largely lost. Overall, GRR illustrates a fundamental limitation of point-wise LDP on rich mobility traces: either we choose $\epsilon$ small enough to have meaningful protection but destroy most analytic value, or we choose $\epsilon$ large enough to preserve utility but provide little effective defense against trajectory-based re-identification attacks. 

\begin{figure}[t!]
  \begin{subfigure}[Home and Work Location Error]{\includegraphics[width=0.91\linewidth]{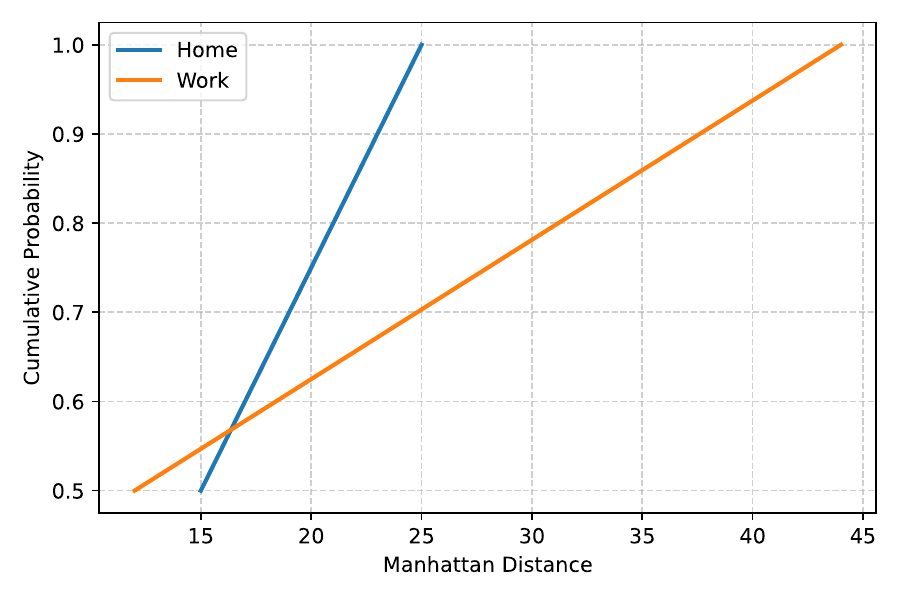}\label{fig:cdf_home_work_error_permuted}}
  \end{subfigure}
  \hfill
  \begin{subfigure}[Utility Loss]{\includegraphics[width=0.91\linewidth]{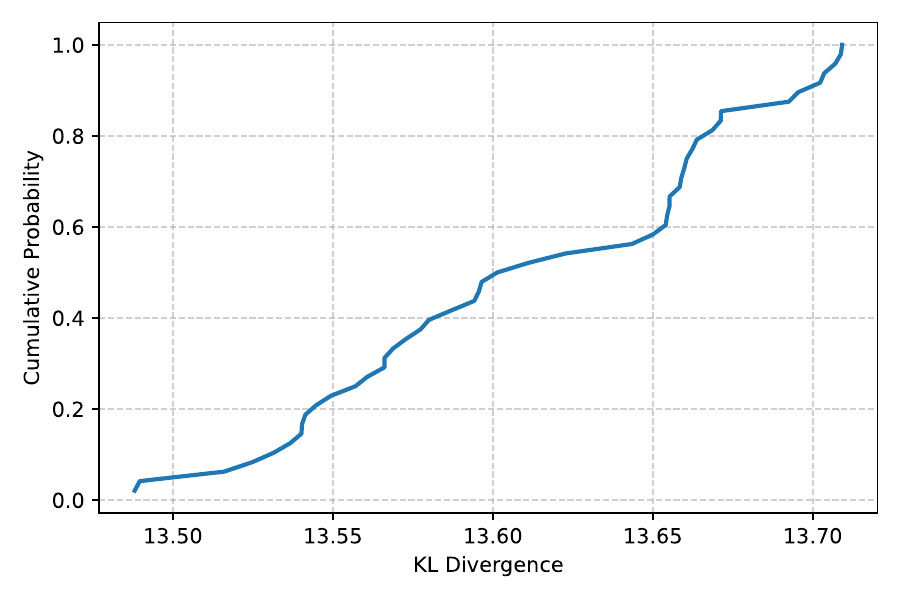}\label{fig:cdf_kl_divergence_permuted}}
  \end{subfigure}
  \vspace{-0.45cm}
  \caption{
    Spatial de-structuring effects.
    (a) Home/work inference error.
    (b) KL divergence after permutation. 
  }
  \label{fig:permutation_metrics}
\end{figure}

\subsection{\texttt{Spatial De-Structuring}}

We finally consider an extreme sanitization mechanism designed to probe how much re-identification power remains when spatial structure is deliberately destroyed. Prior work, including Section~\ref{sec:reidentification-spatial} and~\cite{pinter_revealing_2024}, shows that the \emph{shape} of spatial density: how activity concentrates across space, often uniquely characterizes a city. To counter this effect, we apply a \emph{spatial de-structuring} transformation that randomly permutes the grid-cell indices independently for each user. This permutation preserves, for every cell, its marginal visit frequency and the temporal usage pattern, but breaks the geographic layout: neighbor relationships and large-scale spatial geometry are no longer meaningful. 

\vspace{0.2cm}
\noindent\textbf{Observations:} After de-structuring, the Spearman correlation between clustered user density and census data drops substantially to 40\%, confirming that spatial coherence has largely collapsed. However, coarse density signals remain: Nagoya still emerges as the top candidate city, with a correlation margin exceeding 10\% over any other city under consideration. Figure~\ref{fig:cdf_home_work_error_permuted} shows the CDF of home/work inference error after permutation, revealing a strong degradation in anchor accuracy, while Figure~\ref{fig:cdf_kl_divergence_permuted} reports very large KL divergence between true and estimated population distributions, indicating extreme utility loss. Thus, even when we erase spatial order to an unrealistic degree, residual density patterns still support meaningful city-level inference, but only at the cost of rendering the data nearly unusable for most practical analyses.

\subsection{Takeaways}

Our experiments with various sanitization methodologies collectively paint a stark picture: current mechanisms are fundamentally ill-equipped to protect rich mobility trajectories without inflicting severe damage on utility. 
These findings underscore that trajectory re-identification is not a corner case but a structural vulnerability of mobility data: even after applying state-of-the-art point-wise mechanisms or extreme spatial transformations, our attacks continue to extract meaningful information. In other words, our trajectory re-identification study raises a clear alarm that existing anonymization techniques are misaligned with the complexity of real-world traces. Designing mechanisms that suppress structural patterns: recurrent routes, anchor constellations, and group co-location, while preserving high-resolution spatio-temporal utility, remains an open and pressing research problem. Developing such trajectory-aware privacy protections is, in our view, a necessary next step for any realistic deployment of large-scale mobility datasets. 



\section{Conclusion}

This study demonstrates that current anonymization practices for mobility data are dangerously inadequate. Using \texttt{YJMob100K}~\cite{yabe2024YJMob100K} as a case study, we show that a structure-aware adversary can recover both the underlying geography and real-world timeline, enabling direct re-identification of sensitive patterns such as presence in sparse regions and large-scale uniqueness of home--work anchors. Despite coordinate shifting and timestamp normalization, users remain exposed to location inference, anchor profiling, and downstream linkage through public auxiliary data. These results are a strong warning: as long as structural information in space and time is left intact, “anonymized” trajectories can still be traced back to individuals. There is an urgent need for new privacy mechanisms that explicitly target this structural leakage, and trajectory re-identification should now be treated as a central threat model for the design of future spatio-temporal data releases.

\section*{Responsible Disclosure}

The vulnerabilities uncovered in this paper show that users included in the public \texttt{YJMob100K} dataset face concrete risks of privacy compromise. 
Prior to conducting our analysis, we obtained approval from the ethical and legal committee of our institution. 
Upon confirming our findings, we notified the authors responsible for publishing the \texttt{YJMob100K} dataset and shared a technical summary of the identified vulnerabilities. 
We also disclosed the issues to the relevant Data Protection Authorities, including the Japanese \emph{Personal Information Protection Commission}~\cite{ppc}, to inform them of their potential impact.


\begin{acks}
The authors acknowledge support from Inria (Moyens Incitatifs 2024), ANR IPOP (grant ANR-22-PECY-0002), and ANR GTTP (grant ANR-24-CE25-1133). 
The work of Héber H. Arcolezi was partially supported by the French National Research Agency through ANR AI-PULSE (grant ANR-24-CE23-6239) and the ANR MIAI Cluster (grant ANR-23-IACL-0006).
\end{acks}

\bibliographystyle{ACM-Reference-Format}
\bibliography{references}

\appendix

\section{Events Hosted at the Exhibition Center}\label{events_table}

\begin{table}[H]
\centering
\small
\caption{Events hosted at Port Messe exhibition center between 15/9/2019 and 28/11/2019.}
\label{tab:events_PortMesse}

\begin{threeparttable}
\begin{tabular}{|p{0.43\linewidth}|p{0.33\linewidth}|c|}
\hline
\textbf{Event Name} & \textbf{Dates} & \textbf{Attendance} \\ \hline\hline

\emph{Automotive World Nagoya} 2019\tnote{a} &
September 18--20 (Day 3--5) & unknown \\ \hline

\emph{MECT2019}\tnote{b} &
October 23--26 (Day 38--41) & 90{,}000 \\ \hline

\emph{Messe Nagoya 2019}\tnote{c} &
November 6--9 (Day 52--55) & 60{,}000 \\ \hline

\emph{One OK Rock Concert}\tnote{d} &
November 12--13 (Day 58--59) & unknown \\ \hline

Mammoth Flea Market Z vol.\,61\tnote{e} &
November 16--17 (Day 62--63) & unknown \\ \hline

\emph{Nagoya Motor Show 2019}\tnote{f} &
November 21--24 (Day 67--70) & 205{,}900 \\ \hline

\end{tabular}

\begin{tablenotes}
\footnotesize
\item[a] \url{https://www.automotiveworld-nagoya.jp/en-gb/visit/visitorsguide-dl.html}
\item[b] \url{https://mect-japan.com/2019/en/about/}
\item[c] \url{https://www.nic-nagoya.or.jp/en/nagoya-calendar/assets_c/nagoya_calendar_2019_10.pdf}
\item[d] \url{https://www.japanconcerttickets.com/one-ok-rock-2019-2020/}
\item[e] \url{https://www.nic-nagoya.or.jp/en/nagoya-calendar/assets_c/nagoya_calendar_2019_11.pdf}
\item[f] \url{https://www.jetro.go.jp/en/database/j-messe/tradefair/detail/104782}
\end{tablenotes}

\end{threeparttable}
\end{table}

\section{Sensitive POIs}\label{sensitive_pois}

\begin{table}[H]
  \centering
  \caption{Sensitive POI keyword used to flag cells as sensitive.}
  \label{tab:sensitive_poi_keywords}
  \begin{tabular}{|l|l|}
    \hline
    \textbf{Keyword substring} & \textbf{Keyword substring} \\
    \hline
    Hospital  & Addiction \\
    Clinic    & Rehab \\
    Medical   & Counsel \\
    Pharmacy  & Therapy \\
    Relig     & Politic \\
    Church    & Party \\
    Mosque    & Campaign \\
    Temple    & Adult \\
    Shrine    & Strip \\
    Nightclub &         \\
    \hline
  \end{tabular}
\end{table}

\end{document}